\documentclass[11pt,english]{article}

\usepackage[margin=1in]{geometry}
\usepackage{amsmath,amssymb,amsthm}
\usepackage{xfrac}
\usepackage[dvipsnames]{xcolor}
\usepackage[hyperfootnotes=false, colorlinks=true, linkcolor=black, urlcolor=Black, citecolor={Blue}]{hyperref}
\hypersetup{
    pdfcreator={},   
    pdfproducer={},  
}
\usepackage{framed}
\usepackage[font=small,labelfont=bf]{caption}
\usepackage{tikz}
\usepackage{multirow}
\usetikzlibrary{calc}
\usepackage{csvsimple}
\usepackage{booktabs}
\usepackage{enumitem}

\newcommand{\defined}{\ensuremath{\stackrel{\mathrm{def}}{=}}}

\newcommand{\cQ}{\ensuremath{\mathcal{Q}}}
\newcommand{\cH}{\ensuremath{\mathcal{H}}}
\newcommand{\cM}{\ensuremath{\mathcal{M}}}

\newcommand{\bp}{\ensuremath{p}}
\newcommand{\plotdate}{2021-08-05 - 2021-08-29}

\newcommand{\inparen}[1]{\left( #1 \right) }
\newcommand{\inbrak}[1]{\left[ #1 \right] }

\newcommand{\inset}[1]{\left\{ #1 \right\} }

\newcommand{\inceil}[1]{\left\lceil #1 \right\rceil }

\newcommand{\eps}{\ensuremath{\epsilon}}

\newcommand{\ie}{\emph{i.e.}, }

\newcommand{\suchthat}{\ensuremath{~\middle|~}}

\newcommand{\p}[2]{\ensuremath{#1^{\inparen{#2}}}}
\newcommand{\vset}{\ensuremath{\mathcal{V}}}
\newcommand{\cset}{\ensuremath{\mathcal{C}}}

\newcommand{\score}{\ensuremath{\mathsf{score}}}

\newcommand{\ord}[3]{\ensuremath{{#1}_{#3,#2}}}

\newcommand{\PBPDF}{\ensuremath{\p{f}{PB}}}
\newcommand{\Bin}{\ensuremath{\operatorname{Bin}}}
\newcommand{\PBCDF}{\ensuremath{\p{F}{PB}}}
\newcommand{\Var}{\ensuremath{\operatorname{Var}}}

\newcommand{\du}{\mathop{\dot{\bigcup}}}

\newtheorem{remark}{Remark}
\newtheorem{lemma}{Lemma}
\newtheorem{theorem}{Theorem}

\newtheorem{corollary}{Corollary}
\newtheorem{claim}{Claim}
\newtheorem{definition}{Definition}
\newtheorem{proposition}{Proposition}

\newtheorem{result}{Result}

\renewenvironment{proof}[1][\proofname]{{\bfseries #1.}}{\qed}

\usepackage{natbib}
\bibpunct[, ]{(}{)}{,}{a}{}{,}%

\usepackage{float}

\title{Scaling Blockchains: Can Committee-based Consensus Help?}

\author{Alon Benhaim, Brett Hemenway Falk, Gerry Tsoukalas\thanks{Benhaim: UPenn, Dept. of Mathematics, alonb@sas.upenn.edu; Falk: UPenn, Computer and Information Science Dept., fbrett@cis.upenn.edu; Tsoukalas: Boston University, Questrom School of Business \& The Luohan Academy; gerryt@bu.edu} }

\begin{document}
\pagestyle{plain}
\date{November 22, 2022}

\maketitle

\begin{abstract} 
\footnotesize

In the high-stakes race to develop more scalable blockchains, some  platforms (Binance, Cosmos, EOS, TRON, etc.) have adopted committee-based consensus (CBC) protocols, whereby the blockchain's record-keeping rights are entrusted to a committee of elected block producers. In theory, the smaller the committee, the faster the blockchain can reach consensus and the more it can scale. What's less clear, is whether such protocols ensure that honest committees can be consistently elected, given  blockchain users typically have limited information on who to vote for. We show that the \emph{approval voting} mechanism underlying most CBC protocols is complex and can lead to intractable optimal voting strategies. We empirically characterize some simpler intuitive voting strategies that users tend to resort to in practice and prove that these nonetheless converge to optimality \emph{exponentially} quickly in the number of voters. Exponential convergence ensures that despite its complexity, CBC exhibits robustness and has some efficiency advantages over more popular staked-weighted lottery protocols currently underlying many prominent blockchains such as Ethereum. \\


\end{abstract}

\thispagestyle{empty}
\setcounter{page}{1}

\section{Introduction}

Permissionless blockchains face a challenging problem: How can anonymous/untrusted decentralized agents all agree on a sequence of events, e.g., transactions, or more general state updates? The Bitcoin Whitepaper \citep{nakamoto2008bitcoin} introduced  ``Nakamoto Consensus'', a novel consensus protocol that allowed participants to reach agreement on the state of a distributed database in the absence of trust and stable identities, paving the way to a new form of decentralized money. Once Bitcoin's success highlighted the value of blockchain technology, a high-stakes race ensued to design improved consensus mechanisms that could shore-up Bitcoin's flaws -- most notably low throughput/scalability,  high economic and environmental costs, and delayed transaction finality (we  expand on these in \S\ref{sec:primer}).

One suggested alternative with the potential to address the above is ``Committee-based consensus'', whereby participants  can delegate the chain's record-keeping rights to a relatively small committee. The core idea is that smaller committees can reach consensus more efficiently, albeit, at the cost of less decentralization. Currently, several prominent blockchains including the Binance Smart Chain (BSC), Cosmos-based chains\footnote{Including the Cosmos Hub, crypto.com's chain Cronos, Thorchain, Axelar, Osmosis, Secret and the defunct Terra blockchain among many others.}, Algorand, EOS, and TRON use this approach,%
\footnote{Note, blockchains like EOS, who raised a record-breaking \$4 billion in its Initial Coin Offering in 2017, and Tron, define themselves as ``Delegated Proof of Stake'' systems, but this branding has become tarnished by the criticisms of their technical design \citep{cosmosvseos,XLCB18} and some of their business behavior \citep{justinsun,EOSsuit,Tronarbitration}.  It is important to note that these criticisms do \emph{not} undermine the core ideas of committee-based consensus, as evidenced by the fact that more prominent blockchains (like Cosmos) rely on it as well.}
though, importantly, they differ in \emph{how} the committee members are chosen.

Despite the prevalence of committee-based consensus protocols in practice, they have received relatively little attention in the academic literature so far. In particular, the question of how participants of blockchain systems, who typically have limited and dispersed information, can optimally choose/elect effective and trustworthy committees to maintain the state of the chain is of critical importance, and is not well understood. This work seeks to shed some light on this issue.

Before diving into further details, we first provide a brief overview of some common consensus protocols, comparing single-leader vs. committee-based approaches.

\subsubsection*{Single-leader vs. committee-based consensus protocols}

 Bitcoin's implementation of Nakamoto consensus relies on single-leader consensus protocol termed ``Proof of Work'' (PoW). In PoW, blockchain users can compete with each other by engaging in ``wasteful'' computations, for the chance of being selected ``block leader.'' 
If selected, they have the right to append a block of transactions to the blockchain, and reap any rewards that come with it. The more computing (hash) power participants have, the higher their odds of selection. This incentive structure, in turn, has lead to an arms race to invest in specialized computing hardware (ASICs), which have little value outside of PoW mining.

One desirable feature of PoW's wasteful computation is that it affords Sybil resistance\footnote{A Sybil attack involves creating a large number of pseudonymous accounts in an attempt to seize control of the network.} by imposing a cost to enter the block-producer ``lottery.'' However, less wasteful alternatives exist: Nakamoto consensus can also be implemented using Proof-of-Stake (PoS), and early PoS protocols like Nxt \citep{NXT} were essentially PoS-based versions of Nakamoto consensus, where the chance of being selected as block leader is proportional to one's token stake, rather than one's computing power.

Blockchain platforms are increasingly gravitating towards PoS-based consensus as this mechanism is generally believed to be more scalable and less wasteful than PoW \citep{john2020economic}. Nonetheless, the way PoS is currently implemented on many blockchains maintains the single-leader design philosophy of Bitcoin's original PoW, and it is therefore prone to some of the very same shortcommings. In addition to questions surrounding bandwidth limitations, single-leader protocols generally suffer from lack of ``instant finality'': transactions aren't considered ``final" until several successive blocks have been appended to the chain, meaning, their internal states can only reach eventual consistency. In practice, these chains can and do ``fork'' unpredictably, creating conflicting versions of message history, for instance, when two producers independently produce blocks at around the same time, or when a malicious producer purposely produces multiple conflicting blocks.%
\footnote{It should be noted that forking frequency on PoW and PoS can be quite different in practice. For instance, Ethereum (PoS) sees hundreds of (short-term) forks every day \citep{EthereumUncles}.  By contrast, Bitcoin (PoW) sees fewer than one fork per month \citep{reorgs}.
}

Committee-based consensus protocols have the potential to address some of the issues raised by replacing the ``single-block-producer'' leader model in Nakamoto consensus, with one that relies on the formation of dynamic committees who write blocks with near-instant finality (assuming consensus is reached quickly within the committee itself).%
\footnote{
When the committee is static, the blockchain is said to be ``permissioned.''  In this work, we focus solely on the permissionless setting, where the committee changes dynamically.} 
At its core, (elected) committee-based consensus is rather simple: users continuously vote to elect their preferred block producers to the committee. Keeping the committee size small improves efficiency: increasing throughput, decreasing latency and allowing for member specialization. Unfortunately, a small number of malicious committee members can also undermine the security of the entire blockchain, thus there is a fundamental tension between performance and robustness -- a small committee is extremely efficient but is more centralized, and may compromise security. 

\subsubsection*{Committee-based consensus designs in practice}

There is some variation in the design of committee-based consensus protocols in practice, and in particular, on the question of how committee members should be optimally chosen. We outline below three of the most popular implemented designs in prominent blockchains:

\begin{itemize}
    \item \textbf{Lottery:}
        Algorand committees are not elected by users, but rather chosen randomly via a stake-weighted lottery.
    \item \textbf{Single-choice voting:}
        Most Cosmos-based chains, including the Cosmos Hub, Cronos, Thorchain, Axelar, Osmosis and Secret, use single-choice  elections to elect a committee, i.e., each user can only vote for a single candidate. As do Tron and the Binance Smart Chain.
    \item \textbf{Approval voting:} EOS (which raised a record breaking \$4 Billion in its ICO in 2017) as well as its forks like Telos use a more general voting mechanism termed ``approval voting'' where voters can approve a collection of candidates rather than focusing their voting power on a single one.
\end{itemize}

All of these blockchains (Algorand, BSC, Cosmos, EOS, Tron etc) may differ widely in their features -- they have different tokenomics, different virtual machines and their committees run different consensus algorithms. Nonetheless, the method for \emph{selecting} committees can be largely divorced from the other features of the system, making it amenable to independent study. Of the three designs above, approval voting is the most general. In fact, we later argue that the other two designs can be thought of special (simple) cases of approval voting. We therefore focus the bulk of our attention in this paper on studying approval voting for electing committees in committee-based consensus protocols.

\subsubsection*{Research Questions}

Several interrelated questions follow: First, how should agents vote for their preferred candidates given they only have partial information? Second, how small can the committee be without undermining security? Third, how does committee-based consensus with approval voting compare to the other two common protocols mentioned above?

\subsubsection*{Summary of  Model}

To answer these questions, we develop an approval voting model under partial information. Block producers can be one of two types, either ``honest'' or ``dishonest,'' and the vote succeeds if a $(1-\bp)$ fraction of the elected committee is honest. This is in line with the analysis of most consensus protocols (like those used in Cosmos, Algorand and EOS) where participants are either honest or ``byzantine,'' and the consensus protocol exhibits a strict phase transition when the number of byzantine participants exceeds a given threshold --- usually $2/3$rds.

Token holders are tasked to elect block producers into a committee, but the voters have limited information about the candidates: they receive private signals about the type of each candidate block producer and vote strategically to try and maximize the probability of electing an honest committee. The election process is based on a variation of approval voting, whereby voters approve of a collection of candidates, and the candidates with the most approvals are elected to the committee \citep{BF07}. As we will discuss later, this is fundamentally different from traditional voting schemes, where voting for more than one candidate means splitting your vote.

Assuming the block producer committee uses a traditional consensus protocol to certify blocks, such as Practical Byzantine Fault Tolerance \citep{CL99}, this imposes a strict threshold effect on the committee:  if fewer than $1/3$rd of the committee members are dishonest ($\bp = 1/3$ in our terminology), they cannot disrupt the consensus protocol, but once more than $1/3$rd of the committee members are dishonest, they can completely subvert the committee (which can result in halting transactions, or executing double-spend attacks).

We seek to characterize agent optimal voting strategies under these conditions. 
Further, guided by some stylized facts emerging from our basic empirical observations, we also consider a restriction of the voting strategy space to two simple and intuitive classes: ``threshold voting,'' where voters vote for all candidates whose (conditional) probability of being honest is above a certain threshold (Definition~\ref{def:threshold}), and ``cardinal voting,'' where voters vote for their top $k$ candidates (Definition~\ref{def:cardinal}).

\subsubsection*{Summary of Results}

Even with this relatively simple model, computing the probability of electing an honest committee turns out to be challenging. In Theorem~\ref{thm:success}, we derive this probability in the most general terms, allowing for specialization to various voting strategy classes. We then proceed to examine the  success probabilities for the two, intuitive, voting strategy classes we consider.

We first analyze a special case where there is only a single voter, and show that it is (mathematically) equivalent to a setting in which all voters can credibly (and costlessly) pool their information. Pooling of information is often regarded as a pure hypothetical exercise, but it is worth studying in our setting because voter incentives are aligned, and there is no obvious downside to sharing one's information with others. Under these conditions, we show that the cardinal voting strategy \emph{is} in fact the optimal strategy (Proposition~\ref{prop:single}). But this result breaks down when there is more than one voter, if signals cannot be shared (Proposition~\ref{prop:cardinal_suboptimal}). 

Proposition~\ref{prop:threshold} gives a closed-form solution for the the probability of electing an honest committee when voters follow the threshold strategy. But threshold voting can be suboptimal even when there is just a single voter Proposition~\ref{prop:threshold_suboptimal}). 

Despite the general suboptimality of these simple strategies (other than in the special signal pooling case), we show that the system is surprisingly asymptotically stable. More specifically, regardless of the strategy considered,  and under relatively weak assumptions, the probability of electing an honest committee tends to one, \emph{exponentially fast,} as the number of voters increases (Theorem~\ref{thm:approximate_optimality}). Thus, although the \emph{optimal} voting strategy may be too complex to be realistically achievable, simple, intuitive voting strategies, that token holders tend to use in practice, exhibit very strong robustness.  

Finally, to address the aforementioned tradeoff between efficiency and blockchain security, we compare the approval-voting mechanism for committee selection to two other popular mechanisms: single-choice voting, and lottery selection.  We find that approval voting typically requires much smaller committee sizes (1 to 2 orders of magnitude) to attain the same levels of security (defined as failure tolerance).

Overall, our results suggest that for most practical purposes, committee-based consensus is efficient and robust to the complexity it introduces on the agent strategy space, as long as enough voters are participating in the system. Some limitations are discussed in Section~\ref{sec:fairness}.

In the appendix, Section~\ref{sec:primer}, we discuss in more depth, some of the basics of blockchain consensus protocols and the approval voting mechanism that we model. Readers already familiar with these concepts can skip the section, or its relevant parts, without loss.

\section{Literature Review}

Committee-based consensus is widely used in the blockchain space
\citep{byzcoin,Meng2018,EOS,TRON,cosmosvalidators}, but the academic literature is arguably still lagging behind. Of the few studies we could find, \cite{Meng2018}, \cite{yang2019delegated}, \cite{hu2021improved} examine related topics, but they focus mostly on hypothetical tweaks that could be added to improve existing systems. In contrast, we seek to formally analyze and understand whether the existing systems themselves are robust and efficient, given voters have limited information.

Approval voting was introduced into the blockchain space in Delegated Proof of Stake (DPoS), and the first literature on DPoS started with practitioners, where it was often asserted that DPoS consensus is a more efficient and democratic version of the standard PoS mechanism \citep{binanceDPOS, gemini}. 
The approval voting mechanism underlying DPoS is described in the original whitepaper, \cite{DPoSWP}, but there is little attempt to assess potential agent voting behavior and what could go wrong with it.

Approval voting has been widely studied in the context of political elections \citep{BF07}, and we highlight here some facts about the known dynamics of approval voting in general. In a $k$-winner election system, it is desirable to have the property that if a candidate is ranked first by at least $n/k$ of the voters, then that candidate should be elected to the committee.  Unfortunately, this property does \emph{not} hold under approval voting \citep{EFSS17}.

Similarly, an approval voting scheme can end up electing candidates that would lose a majority of pairwise contests against the other candidates, i.e., an approval voting scheme may elect a ``Condorcet loser'' \citep{N84}.

One of the most interesting features of approval voting schemes is that voters typically have multiple honest strategies \citep{N84}. For example, consider an up-to-$2$, $2$-winner system with two voters ($n=2$), and four candidates ($m=4$), $\cset = \{c_1,c_2,c_3,c_4\}$. If the two voters’ preference orders are $(c_2; c_3; c_4; c_1)$ for voter $1$ and $(c_1; c_3; c_4; c_2)$ for voter $2$ then the candidate $c_3$ will be in the elected committee for any $t \geq 2$, so both $c_1$ and $c_2$ cannot be in the elected committee. Should voter $1$ vote for $c_2$ only, or $c_2$ and $c_3$?  These are both honest strategies, and thus even honest players must think strategically. This feature makes the analysis of approval voting systems complex.

Committee-based consensus mechanisms that use approval voting inherit these aforementioned properties, but differ from traditional approval voting in several ways that we describe in the model section. The most significant departure is perhaps that extant studies (outside of the blockchain literature) assume voters have competing interests, and usually have perfect information about the candidates themselves. By contrast, in the blockchain protocol setting,  voter interests can be more aligned.  All voters wish to elect an honest committee, but they have limited information about the candidates. This completely changes the nature of the analysis.

Though we are not aware of any studies considering strategic agent voting behavior in committee-based protocols, numerous studies have looked at strategic agent behavior in other Blockchain protocols. \cite{saleh2021blockchain, rocsu2021evolution, fanti2019economics} are some of the first studies looking at the economics of PoS systems. \cite{LSRDPG20} study weighted voting in validator committees in PoS protocols. There is also a relatively large computer science literature blending strategic considerations and technical design elements of PoS, such as \cite{gavzi2019proof, algorand, bentov2016snow, kiayias2017ouroboros}.

Beyond PoS, \cite{alsabah2020pitfalls, biais2019blockchain, cong2021decentralized, garratt2020fixed} focus on the economics of PoW, and the underlying mining mechanism. Several other studies focus more specifically on Bitcoin, such as \cite{nakamoto2008bitcoin, easley2019mining, huberman, pagnotta, prat2021equilibrium}.

    Other works have considered consensus in the presence of three types of participants byzantine, altruistic and rational \cite{BAR}, or just byzantine and rational \cite{ABPP20}.  In the byzantine-rational model of consensus \cite{ABPP20}, there are still two types of participants, and there is still a phase transition when the number of byzantine participants exceeds a certain threshold, thus our analyses applies almost equally in this setting as well.

Finally, on a broader note, our work is related to the literature studying security guarantees for different types of blockchain protocols, e.g., \cite{lewis2020general, lewis2021does}, though we are not aware of any prior work focused specifically on committee-based consensus. More generally,  our work also has implications for the literature studying the economics of token systems, see e.g., \cite{cong2021tokenomics, tsoukalas2020token, gan2021initial, gan2021infinity}.

To the best of our knowledge, ours is the first paper to analyze the efficiency of committee elections in committee-based consensus protocols, with private information and strategic voters.

\section{Preliminaries and Empirical Observations}

\subsection{Definitions}
Approval voting is a system where each voter may select (``approve") any number of candidates, and the winners are the candidates approved by the largest number of voters (see \cite{K10} for a survey on approval voting). Formally:

\begin{definition}[$k$-winner Approval Voting]
	\label{def:kwinner}
	A set of voters $\vset$ votes on a set of candidates, $\cset$.  Let $n \defined |\vset|$, and $m \defined |\cset|$.
	Voter $v$ chooses a subset of candidates $\cset_v\subseteq \cset$ they wish to vote for. For each candidate $c \in \cset$, the \emph{score} of candidate $c$, that is, the number of votes the candidate receives, is defined to be
	\begin{equation}
		\score(c) \defined \left| \inset{ v \suchthat c \in \cset_v} \right|.
	\end{equation}
	The elected committee is determined to be the $k$ candidates with the highest scores.
\end{definition}

In blockchain settings, stake-holders typically vote for a set of ``block-producers''
modifying the  \emph{k-winner Approval Voting} system to  include a cap on the number of candidates. Formally: 

\begin{definition}[up-to-$t$-vote, $k$-winner Approval Voting]
	\label{def:tkwinner}
	With notation as in definition \ref{def:kwinner}, we limit the maximum number of candidates $t$ each voter can vote for, so that voter $v$ chooses a subset of candidates $\cset_v\subseteq \cset$ restricted to $|\cset_v|\leq t(\leq m)$. As before, the elected committee is determined to be the $k$ candidates with the highest scores.
\end{definition}

    We will assume throughout that there are at least $k$ candidates, $m\geq k$. In general, there may be less than $k$ candidates in the elected committee if less than $k$ candidates received any votes. Alternatively, there may be more than $k$ candidates if there are ties. We specify how we handle these cases in Definition \ref{def:hcommittee}.

\subsection{Empirical Observations: Approval Voting on EOS}
\label{sec:eos}
	Block Producers on EOS are elected by token holders according to a up-to-$30$-vote, $21$-winner approval voting system (see Definition~\ref{def:tkwinner}, with $t=30$ and $k=21$). The $21$ winning candidates form the block producer committee.  Elections are held continuously, and each committee of block producers remains in control of the chain for 126 seconds \citep{EVG18}.
	
	EOS voters are not directly rewarded for staking (although this has been proposed as in \cite{ENY19}), instead voters are assumed to benefit indirectly from the stability and performance of the platform. In EOS and other DPoS systems, votes are weighted by stake, and voters are allowed to ``proxy'' their votes, i.e., delegate their voting power to a different voter.  
	
	To understand the actual voting strategies employed by users, we extracted voting data from EOS. As the EOS blockchain is extremely large (over 8TB) and the majority of transactions are unrelated to voting, we gathered daily voting snapshots from EOS Authority (a block producer) and we used these to analyze voter behavior during the period \plotdate \ (the findings are consistent and exhibit relatively little variability, over different/longer time windows).  Each snapshot contained the current votes of the nearly 1 million accounts that have ever voted.
	
\begin{figure}[H]
	\begin{center}
	    
		\includegraphics[width=.48\textwidth]{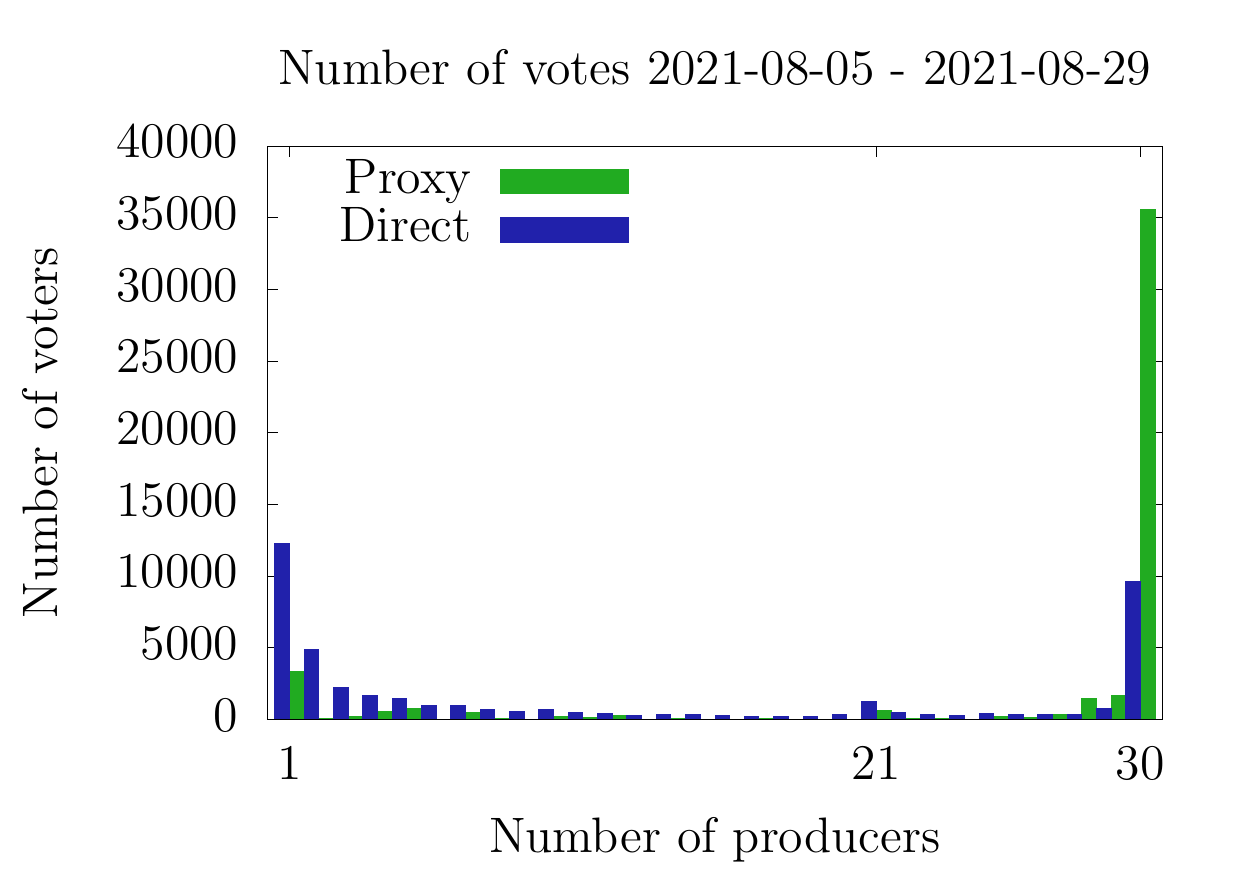}
		\includegraphics[width=.48\textwidth]{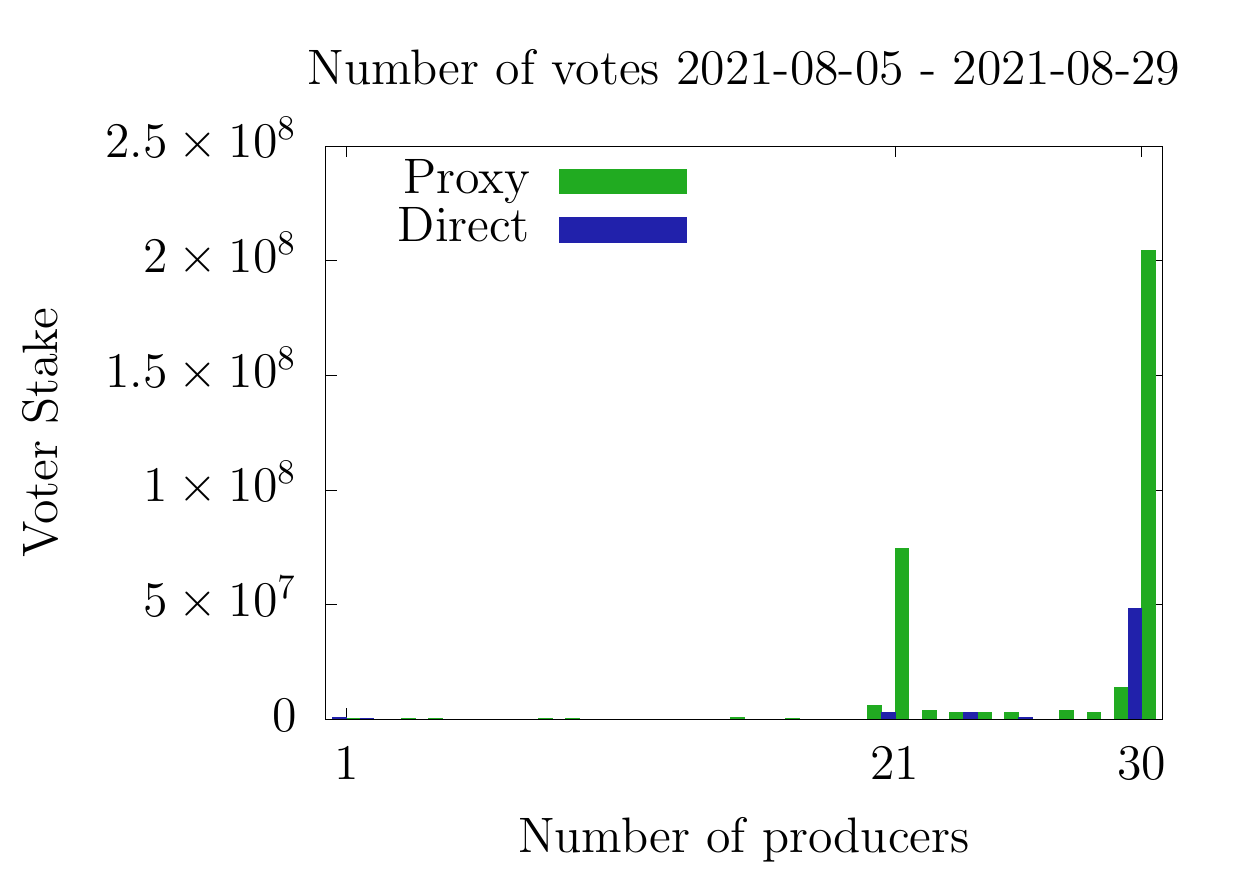}
		\caption{The number of producers that each token holder voted for, during the period \plotdate.  Left panel: unweighted voting. Right panel: stake-weighted voting. \emph{Key Takeaway:} most voters follow a ``cardinal voting'' strategy.
		\label{fig:voter_producers}}
	\end{center}
\end{figure}

	Figure~\ref{fig:voter_producers} shows the number of votes cast by individual voters (left panel), and stake-weighted votes (right panel), on a typical day. 
	
    The key observation is that {\bf most voters follow a ``cardinal voting strategy''} where they vote for a fixed number of block producers, either 1, 21 or 30. A formal definition of ``cardinal voting'' is given in Definition~\ref{def:cardinal}. 
    
    We rely on these basic empirical observations to inform our model in Section~\ref{sec:voting_model}. In Section~\ref{sec:results}, we explore the optimality of these types of intuitive voting strategies.

\section{Model} \label{sec:voting_model}

\subsection{Approval Voting with Limited Information}
We lay out a simple model, where the blockchain's token holders vote according to an up-to-$t$, $k$-winner approval voting system (Definition~\ref{def:tkwinner}), to elect 
a committee of block producers.  There is a pool of $m$ candidates to choose from, $c_1,\ldots,c_m$, and $n$ strategic voters on the platform, $v_1,\ldots,v_n$. Every producer has an unknown type, either ``honest'', $H$, or ``malicious'', $M$. The goal of each voter is to maximize the probability that a supermajority (e.g. a $(1-\bp)$-majority) of the elected committee is honest. We discuss some possible alternative objectives in Appendix~\ref{sec:quality}.

\begin{definition}[Honest Committee]
	\label{def:hcommittee}
	Suppose the $k$ producers with highest number of votes are elected to be on the block-producer committee, $\mathbb{T}$. If there are less than $k$ candidates with non-zero score, then the committee is filled adversarially (i.e., in a worst-case fashion), and if there are ties between the candidates such that there are more than $k$ producers with highest score, then they are broken adversarially (between the ones with least score). Since most Byzantine Agreement protocols require at least $\inceil{\inparen{1-\bp}\cdot k}$ honest members, we say the committee $\mathbb{T}$ is \emph{honest}, $\mathbb{T}=H$, if at least $\inceil{\inparen{1-\bp} \cdot k}$ of the elected block producers are honest.
\end{definition}

Suppose the \textit{a priori} probability that producer $j$ is honest is $\Pr[c_j=H]=p_j$. Also, suppose voter $v_i$ receives a private noisy signal vector, $\mathbf{s}^*_{i}=(s^*_{ij})_{j=1}^m$ about producer's $c_j$ honesty,
\begin{equation}
	\label{eqn:signal}
	s^*_{ij} = \left\{ \begin{array}{l} p_h + \eps_{ij} \mbox{ if Producer $j$ is honest} \\ p_m + \eps_{ij} \mbox{ if Producer $j$ is malicious, } \end{array} \right.
\end{equation}
where $\eps_{ij}$ is a normally distributed noise term with $\mathbb{E}[\eps_{ij}]=0$ and $\Var[\eps_{ij}] = \sigma^2_{ij}$, i.e., $\eps_{ij} \sim \mathcal{N} (0,\sigma^2_{ij})$,$\forall j \in \{1,\ldots,m\}$. It follows that signals are normally distributed with 
$s^*_{ij} \sim \mathcal{N} (p_h,\sigma^2_{ij})$ if producer $j$ is honest, and $s^*_{ij} \sim \mathcal{N} (p_m,\sigma^2_{ij})$ if producer $j$ is malicious. 

We assume that $(p_j)_{1\leq j \leq m},(\sigma_{ij})_{1\leq i\leq  n,1\leq j \leq m},p_h,p_m$, are publicly visible, but voters cannot observe others' private signals.

When voter $i$ receives a signal, $s_{ij}^*$, regarding producer $j$, the voter can compute the \emph{posterior} probability that producer $j$ is honest \emph{conditioned on} $s_{ij}^*$:

\begin{equation}
    s_{ij} \defined \Pr \inbrak{ \mbox{ producer $j$ is honest } \suchthat s_{ij}^* }.
\end{equation}

The map $s_{ij} \leftrightarrow s_{ij}^*$ is a bijective function, and we calculate it explicitly in Lemma~\ref{lem:posterior} in Appendix~\ref{app:posterior}. The result is given below in \eqref{eq:posterior1}.
\begin{equation}\label{eq:posterior1}
s_{ij} =\frac{ 1 }{1 + \frac{1-p_j}{p_j}e^{\frac{(s^*_{ij}-p_h)^2 - (s^*_{ij}-p_m)^2}{2 {\sigma_{ij}}^2} } }.
\end{equation}

After observing their private signal, voters simultaneously submit their ``votes" $\cset_{v_i} = \cset_{v_i}(\mathbf{s}_i)$; we consider any voting strategy in the class of voting strategies, represented by the letter $\mathbb{S}$.

Voter $i$'s payoff is given by $u_i$ if the elected committee is honest ($\mathbb{T} = H$) and $0$ otherwise.
We assume voting has some unit cost, $c$ (the opportunity cost of staking one unit of capital).

Voter $i$'s objective is to maximize their utility given by
\begin{equation}
    u_i \Pr\inbrak{ \mathbb{T} = H } - c.
\end{equation}

With exogenous $c$ and $u_i$, this objective simplifies to maximizing the success probability $\Pr\inbrak{ \mathbb{T} = H }$ --- the probability that the elected committee $\mathbb{T}$ is honest, conditioned on the private signal vector a user receives, $\mathbf{s}_i$, given the platform voting system in Definition \ref{def:tkwinner}. Formally:

\begin{equation}
    \label{eq:max_prob}
		\max_{\cset_{v_i}\subseteq \cset} {\Pr \inbrak{\mathbb{T}=H \suchthat \inset{ \mathbf{s}_i}_{i=1}^n}}.
\end{equation}

\begin{definition}[Voting Strategy]
    \label{def:voting_strategy}
	A voting strategy is an algorithm $A$ used by voters, that takes as input the parameters the voter has access to $\mathbf{s}_{i}=(s_{ij})_{j=1}^{m},(p_j)_{j=1}^{m},(\sigma_{ij})_{j=1}^{m},p_h,p_m$ and outputs a subset of candidates the voter wishes to vote for $\cset_v\subseteq \cset$. We denote the committee elected by exerting algorithm $A$ as $\mathbb{T}_A$.
\end{definition}

 With Definition ~\ref{def:voting_strategy}, we can interchangeably talk about the voters maximizing the success probability by exerting a voting algorithm $A\in \mathbb{S}$ and rewrite Equation~\ref{eq:max_prob} as:
\begin{equation}
    \label{eq:max_prob_alg}
		\max_{A\in \mathbb{S}} {\Pr \inbrak{\mathbb{T}_A=H}}.
\end{equation}

We say that a strategy is optimal if it maximizes the success probability - the probability of electing an honest committee, Equation~\ref{eq:max_prob} or \ref{eq:max_prob_alg}. Table~\ref{tab:notation} summarizes the notation.

\begin{table}[h]\small
    \centering
    \begin{tabular}{|l|l|}
        \hline
        $m$ & Number of (candidate) block producers\\
        \hline
        $n$ & Number of voters \\
        \hline
        $p_j$ & \textit{A priori} probability block producer candidate $j$ is honest\\
        \hline
        $p_m$ & The base signal for a malicious candidate producer\\
        \hline 
        $p_h$ & The base signal for an honest candidate producer\\
        \hline
        $\sigma_{ij}$ & Standard deviation of the noise, $\eps_{ij}$ for voter $i$, and producer $j$\\ 
        \hline
        $k$ & Elected Committee size \\
        \hline
        $s_{ij}^*$ & Voter $i$'s raw signal about producer $j$ \\
        \hline
        $s_{ij}$ & Producer $j$'s posterior probability of being honest, conditioned on $s_{ij}^*$. \\
        \hline
    \end{tabular}
    \caption{Notation}
    \label{tab:notation}
\end{table}

While voters' optimization problem is well-defined, computing the objective function $\Pr \inbrak{\mathbb{T}_A=H}$ is challenging. As a first step, we need to define the types of voting strategies that are accessible to agents. This is the objective of the next section.

\subsection{Class of Voting Strategies}

In principle, any function $f: [0,1]^m \rightarrow \inset{0,1}^m$ is a possible voting strategy.  It seems clear, however, that any reasonable strategy should be coordinate-wise non-decreasing, i.e., if $s_j' > s_j$ and $f(s_1,\ldots,s_{m}) = (y_1,\ldots,y_m) \subset \inset{0,1}^m$, and if $f(s_1,\ldots, s_{j-1}, s_j', s_{j+1}, \ldots,s_m)= (y_1',\ldots,y_m')$ then $y'_j \ge y_j$.  In other words, if one candidate's signal increases (while the other signals remain the same) this \emph{cannot} cause the voter to switch their vote away from the candidate. We refer to this class as the ``general class'' of voting strategies (previously referred to as class $\mathbb{S}$).

Within this this general class, we also consider in our analysis  two particularly simple and intuitive strategies related to our empirical observations: \emph{threshold voting} (Definition~\ref{def:threshold}) and \emph{cardinal voting} (Definition~\ref{def:cardinal}). 

\begin{definition}[Threshold Voting]
	\label{def:threshold}
	Voter $v_i$ is said to follow the \emph{``threshold'' voting strategy} if they choose a threshold $z_i \in [0,1]$ and vote for all producers $c_j$, with posterior probability higher than this threshold, formally: $s_{ij}=\Pr \inbrak{c_j = H \suchthat s^*_{ij}} > z_i$.
\end{definition}

If we define $p_{ij} \defined \Pr \inbrak{ s_{ij} > z_i }$,
then $p_{ij}$ is the probability that voter $i$ votes for producer $j$ (assuming voter $i$ is following the threshold voting strategy).  Summing over all $n$ voters, the number of votes received by producer $j$ is distributed as the sum of $n$ Bernoulli random variables with parameters $p_{1j},\ldots,p_{nj}$.  If $p_{1j} = \cdots = p_{nj}$, then the number of votes received by producer $j$ is a binomial random variable.  When the $p_{ij}$ are distinct, then the number of votes received by producer $j$ is a \emph{Poisson Binomial Random Variable}. See Appendix~\ref{app:PBD} for a review of useful properties of the Poisson Binomial Distribution.

This characterization of the distribution of votes when voters follow the \emph{threshold voting strategy} will be important as we study the dynamics of this strategy in Section~\ref{sec:results}.

\begin{definition}[Cardinal Voting]
	\label{def:cardinal}
	Voter $v_i$ is said to follow the \emph{``cardinal'' voting strategy} if they choose to vote for the top $z_i \in \inset{1,\ldots,t}$ producers with the highest posterior probabilities of being honest $s_{ij}=\Pr \inbrak{c_j = H \suchthat s^*_{ij}}$.
\end{definition}

When voters follow the \emph{cardinal voting strategy}, the number of votes received by each producer is still distributed as a Poisson Binomial random variable, but now the parameters $p_{ij}$ (the probability that voter $i$ votes for candidate $j$) are much more challenging to compute, as we will discuss in the analysis.

Connecting this back to the empirical voting strategies discussed in Section~\ref{sec:eos}, Figure~\ref{fig:voter_producers} shows that EOS voters tend to follow the cardinal voting strategy with $z = 1, 21$ or $30$.

\section{Analysis} \label{sec:results}

We begin our analysis by characterizing the probability of electing an honest committee in the most general terms possible (Section~\ref{sec:probhonest}) and unveiling some associated complexities (Section~\ref{sec:complexity}). We then examine outcomes in a simplified single-voter/signal pooling setting which helps build intuition (Section~\ref{sec:single_voter}), before looking at the general multi-voter case (Section~\ref{sec:multiple_voters}). The preliminary results obtained hint at possible asymptotic optimality, which is formally analyzed in Section~\ref{sec:asymptotic}. Finally, we compare approval voting committee-based consensus to other PoS-based mechanisms (Section~\ref{sec:PoScomparison}).

\subsection{The Probability of Electing an Honest Committee}\label{sec:probhonest}

The voters' goal is to vote in such a way that the probability of electing an honest committee (Definition~\ref{def:voting_strategy}) is maximized.  Before we can maximize this probability, however, we must calculate the probability of electing an honest committee for \emph{a fixed set of strategies}. Computing this probability is quite complex, and we break it down into a series of manageable steps. For exposition, we also focus primarily in the text on \emph{threshold} voting, and leave the \emph{cardinal} voting analysis for the Appendix.

\begin{enumerate}
    \item \textbf{Section~\ref{sec:probvote}: Probability that a user votes for an honest candidate:} First, we calculate the probability that an honest candidate receives a vote from voter $i$.  Proposition~\ref{prop:threshold_suboptimal} calculates this probability when voters follow the \emph{threshold} voting strategy, and Proposition~\ref{prop:cardinal} calculates this probability when voters follow the \emph{cardinal} voting strategy.  
    \item \textbf{Section~\ref{sec:distvote}: Density function of votes received for honest candidates:} Once we have the probability that an honest candidate receives a vote from a given candidate, the \emph{number} of votes received by a given candidate becomes a Poisson Binomial (Theorem~\ref{thm:voting_distribution}).  
    \item \textbf{Section~\ref{sec:honestcommittee}: (Aggregate) Probability of selecting an honest committee:} Once we have the density function of votes for both honest and dishonest producers, we can calculate the probability that a $\bp$-fraction of the elected committee is honest.  Theorem~\ref{thm:success} calculates this probability in closed-form, when voters follow the \emph{threshold} voting strategy. In Appendix~\ref{app:cardinal}, we outline how to adapt Theorem~\ref{thm:success} to the cardinal-voting setting.
\end{enumerate}

\begin{remark}[Voter precision]
Throughout this section, we assume that producers are indistinguishable except for their type, meaning, the variance of the noise $\sigma_{ij} = \sigma_i$, for $1\leq j \leq m$ in Equation~\ref{eqn:signal}. This implies there is a single pdf, $f^h(x)$ that denotes the probability an honest producer receives $x$ votes. This simplification is done purely for expositional purposes; to obtain the result for the more general case, one would need to replace Theorem~\ref{thm:order_cdf} (used in the proof of Theorem~\ref{thm:success}) by the more general Bapat-Beg Theorem \citep{bapatbeg}. The resulting expression remains closed-form, but is too cumbersome for display.
\end{remark}

\subsubsection{Probability of voting for an honest candidate}
\label{sec:probvote}

When voters follow the \emph{threshold} or \emph{cardinal} voting strategy, we can compute the probability $p_i^h$ (resp. $p_i^m$) which is the probability that an honest (reps. malicious) candidate receives a vote from voter $i$.  Proposition~\ref{prop:threshold} gives the probabilities when voters follow the threshold voting strategy.

\begin{proposition}[Threshold voting]
    \label{prop:threshold}
    For a producer $j$, let $p_i^h$ (resp. $p_i^m$) denote the probability that voter $i$ casts a vote for producer $j$ conditioned on producer $j$ being honest (resp. dishonest).
    
	When voters follow the threshold strategy (Definition~\ref{def:threshold}) with threshold, $z_i$,
    \begin{align}
        p_i^h = 1 - \Phi\inparen{ \frac{h^{-1}(z_i) - p_h }{\sigma_i} }, \quad 
        p_i^m = 1 - \Phi\inparen{ \frac{h^{-1}(z_i) - p_m }{\sigma_i} },
    \end{align}
    where $\Phi$ is the density function of the standard normal distribution, and 
    \begin{equation}
        	h^{-1}(q) \defined \frac{ p_h^2 - p_m^2 - 2 \sigma^2 \log \inparen{ \frac{p(1-q)}{(1-p)q} } }{2(p_h-p_m)}
    \end{equation}
    is derived in Lemma~\ref{lem:posterior}.
\end{proposition}

In Proposition~\ref{prop:cardinal} in Appendix~\ref{app:cardinal} we show how to calculate those probabilities for the cardinal-voting strategy.

\subsubsection{Distribution of votes}
\label{sec:distvote}

Propositions~\ref{prop:threshold} and \ref{prop:cardinal} give the probabilities $p_i^h$ and $p_i^m$ that voter $i$ votes for honest, or dishonest candidates.

Note that for a given candidate, $j$, once we condition on the candidate's type (either honest or dishonest), the events that voter $i$ and voter $i'$ vote for candidate $j$ become independent. This means that the \emph{number} of votes received by each type of candidate is distributed according as a Poisson Binomial Random variable.  Theorem~\ref{thm:voting_distribution} gives the general form of the distribution of votes ($f^h,f^m$) received by honest and dishonest producers in terms of the probabilities ($p_i^h$,$p_i^m)$ that voter $i$ casts a vote for producer $j$.  

\begin{theorem}[Distribution of Votes]
    \label{thm:voting_distribution}
    For a producer, $j$, let $p_i^h$ (resp. $p_i^m$) denote the probability that voter $i$ casts a vote for producer $j$ conditioned on producer $j$ being honest (resp. dishonest).
    Then the probability distribution of the number of votes received for honest and dishonest producers is given by 
    \begin{align}
    f^h(x) &= \sum_{A\in F_{x}}{\prod_{i_1\in A}{p^h_{i_1}}\prod_{{i_2}\in A^c}{1-p^h_{i_2}}} \\
    f^m(x) &= \sum_{A\in F_{x}}{\prod_{i_1\in A}{p^m_{i_1}}\prod_{{i_2}\in A^c}{1-p^m_{i_2}}}, 
    \end{align}
    where $F_{x}$ is the set of all subsets of $x$ integers that can be selected from $\inset{1,2,3,...,n}$.
\end{theorem}

\subsubsection{Probability of electing an honest committee}
\label{sec:honestcommittee}

The final step in this analysis is to compute the probability that a $\bp$-fraction of the committee is honest. Notice that when voters follow the \emph{threshold} voting strategy, for a given voter, $i$, and distinct candidates $j$ and $j'$, the probability that $i$ votes for $j$ and $j'$ are independent events. This allows us to prove Theorem~\ref{thm:success}, which gives an analytic representation of the probability of electing an honest committee in the \emph{threshold} voting setting.

\begin{theorem}[Success Probability (threshold voting)]
	\label{thm:success}

	Suppose there are $m$ producers, and each producer is honest independently with probability $p$. If the number of votes received by each candidate are independent random variables then the probability that there are at least $\inceil{\inparen{1-\bp} \cdot k}$ honest producers in a committee of size $k$, is given by:	
	\begin{align}
		\Pr \inbrak{ \mathbb{T} = H } 	&= \sum_{a=m-\inceil{\bp \cdot k}+1}^{m}{ \binom{m}{a} p^{a} (1-p)^{m -a}} + \sum_{a=\inceil{(1-\bp) \cdot k}}^{m-\inceil{\bp \cdot k}} \binom{m}{a} p^{a} (1-p)^{m -a} \sum_{x=0}^n \Biggl[ \nonumber \\
		& \left( \sum_{j=0}^{\inceil{(1-\bp) \cdot k}-1} \binom{a}{j} \left( {\left( 1 - F^h(x) \right)}^j \inparen{F^h(x)}^{a-j} - \inparen{ 1 - F^h(x) + f^h(x) }^j \inparen{ F^h(x) - f^h(x) }^{a-j} \right) \right) \Biggr. \nonumber\\
										&\Biggl. \left( \sum_{j=0}^{\inceil{\bp \cdot k}-1} \binom{b}{j} \inparen{ 1- F^m(x) +f^m(x) }^j \inparen{F^m(x)-f^m(x)}^{b-j} \right) \Biggr], \label{eqn:success}
	\end{align}
	where $f^h,F^h$ are the PDF and CDF of the number of votes received by an honest producer, 
	and $f^m,F^m$ are the PDF and CDF of the number of votes received by a dishonest producer.
	
\end{theorem}

In the \emph{cardinal} voting setting, it is possible to prove an analog of Theorem~\ref{thm:success}.  The exact formula is significantly more complicated, however, because the events that voter $i$ votes for candidate $j$ and candidate $j'$ are no longer independent (they are now negatively correlated). In Appendix~\ref{app:cardinal}, we outline how to adapt Theorem~\ref{thm:success} to the cardinal-voting setting.

Combining Theorems~\ref{thm:voting_distribution} and \ref{thm:success} gives a closed-form expression for the success probability (whenever $p^h_i$ and $p^m_i$ can be calculated).

\subsection{The Complexity of Approval Voting}\label{sec:complexity}

We focus here (without loss) on \emph{threshold} voting. Similar results hold in case of \emph{cardinal} voting.

\begin{figure}
	\begin{center}
		\begin{tikzpicture}
			\node (A) {\includegraphics[width=.45\textwidth]{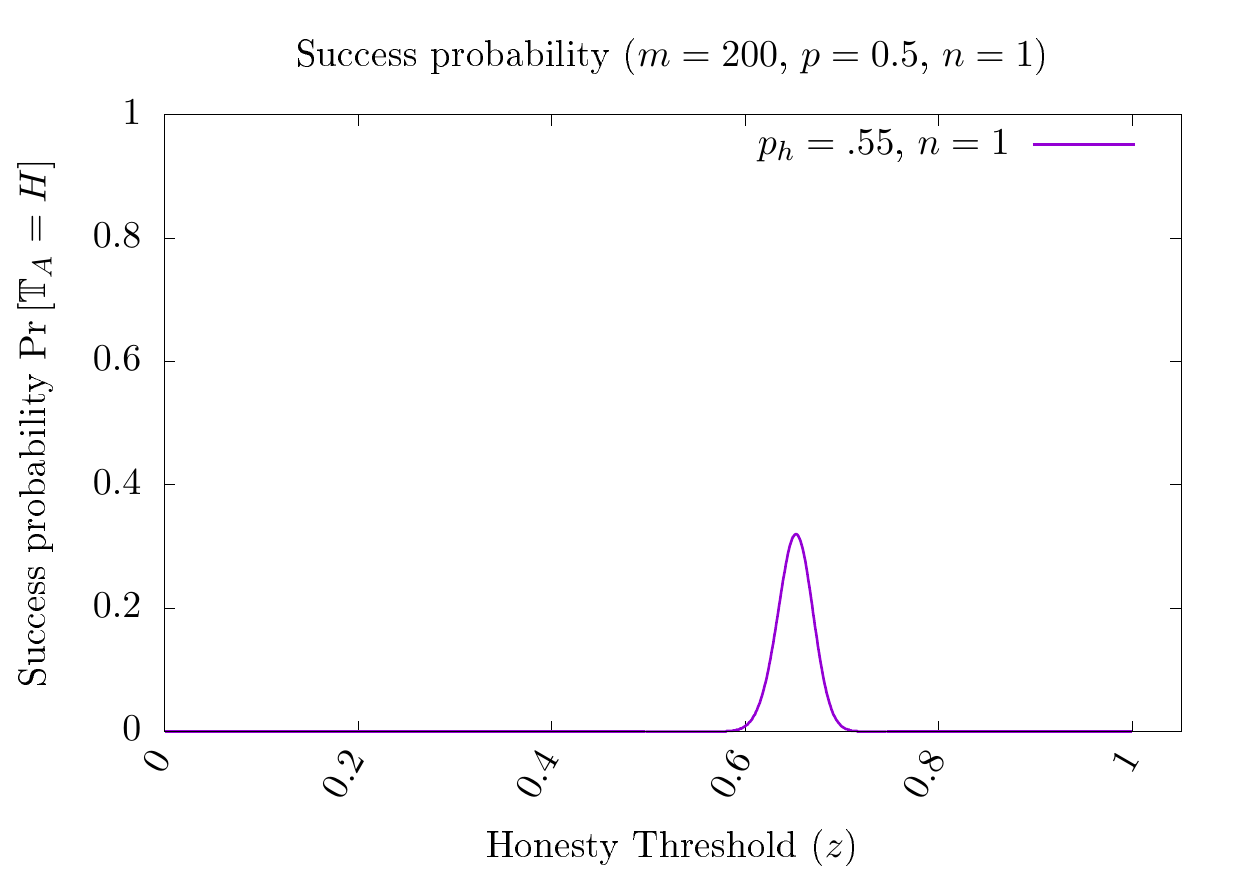}};
			\node (B) at ([yshift=-1cm]A.south) [anchor=north] {\includegraphics[width=.45\textwidth]{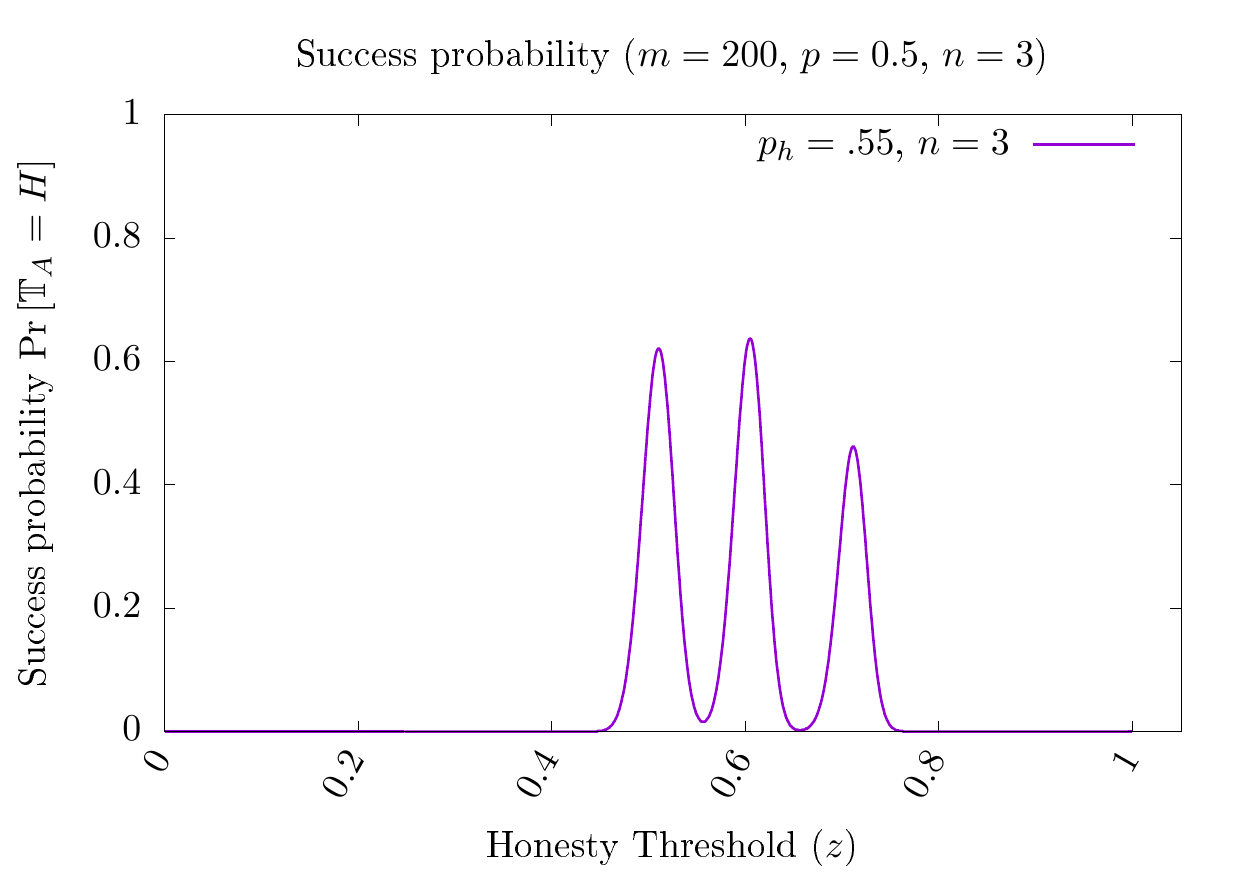}};
			\node (C) at ([xshift=1cm]A.east) [anchor=west] {\includegraphics[width=.45\textwidth]{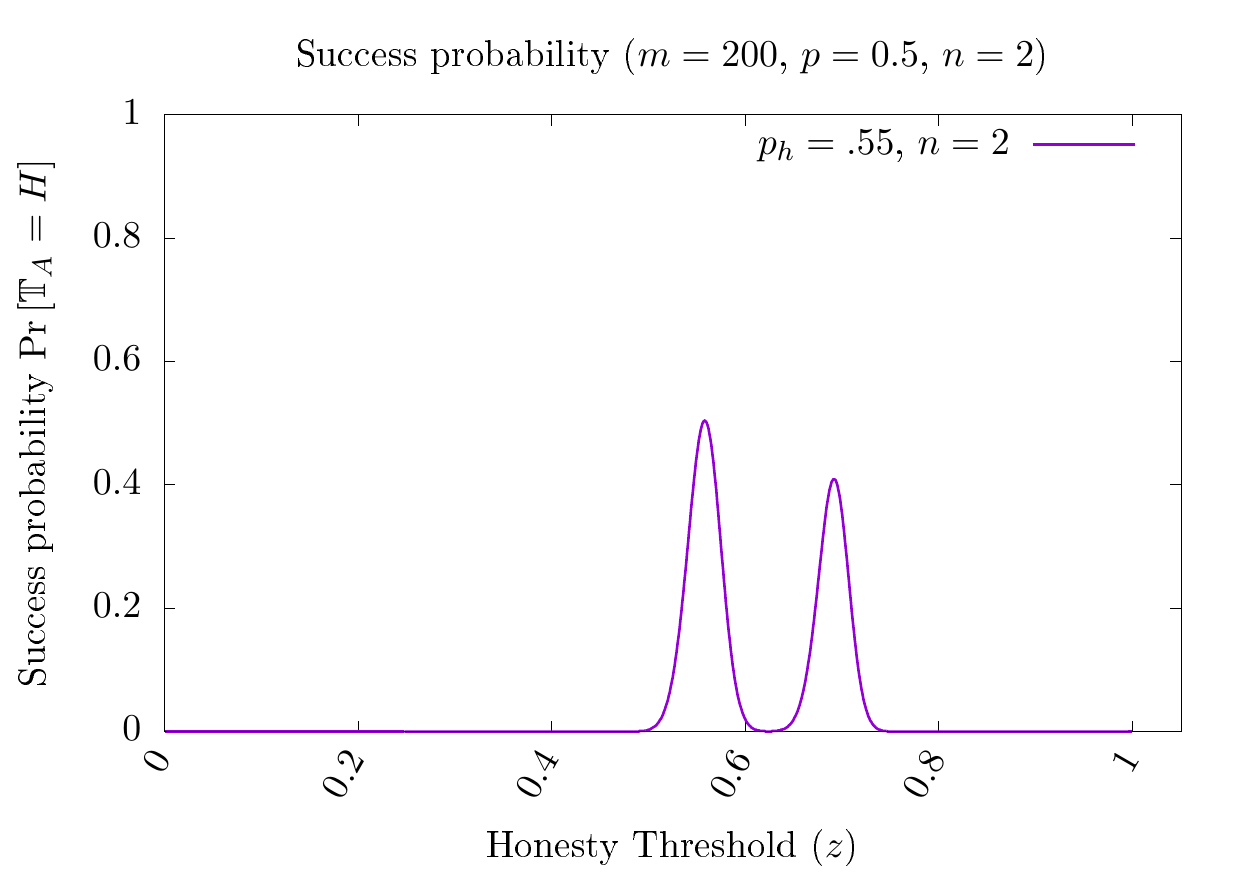}};
			\node (D) at ([xshift=1cm]B.east) [anchor=west] {\includegraphics[width=.45\textwidth]{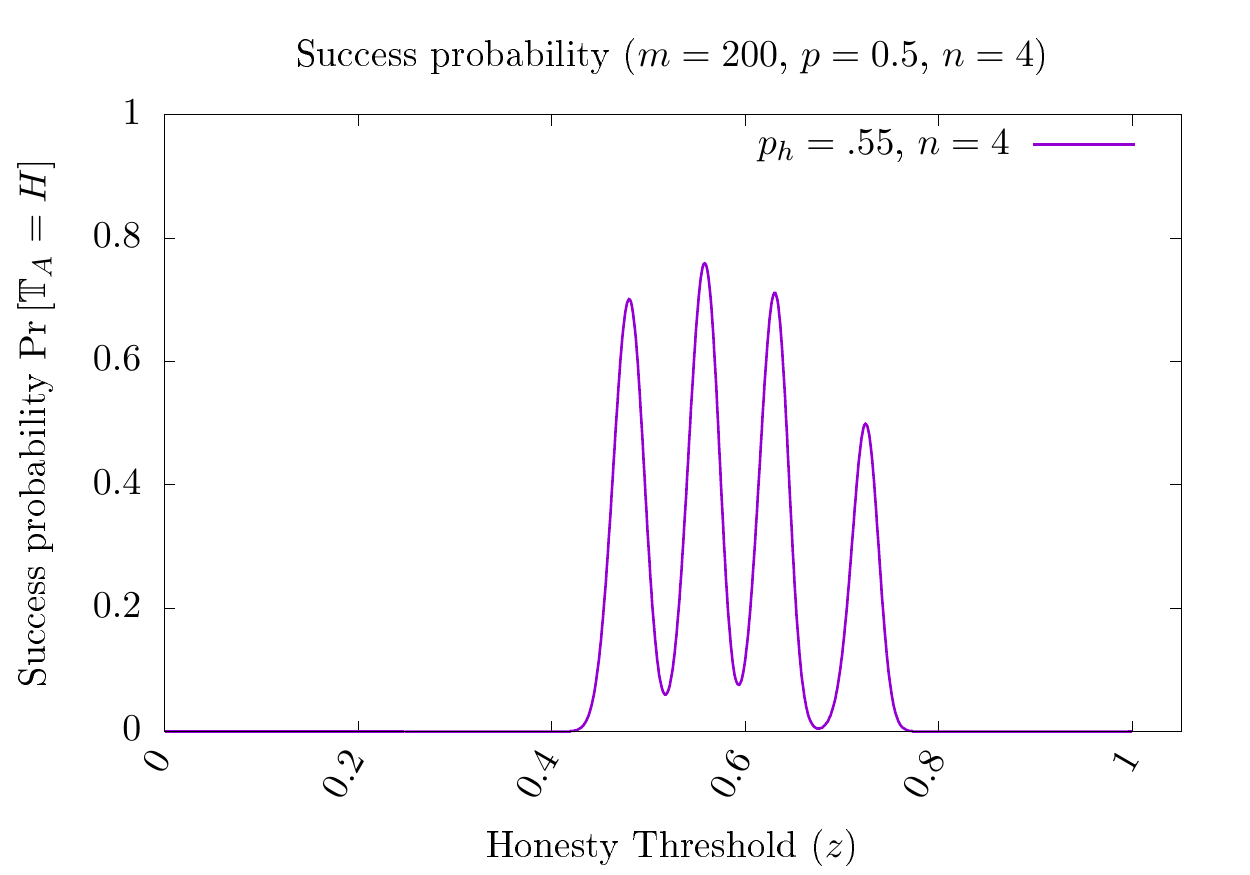}};
		\end{tikzpicture}
		\caption{Success probability as a function of threshold chosen, assuming small number of voters ($n \in \{1, 2, 3, 4\}$), and all voters use the same threshold. \emph{Key Takeway:} The number of local optima increases with $n$. \label{fig:threshold_few_voters}} 
	\end{center}
\end{figure}

Finding the \emph{optimal} voting strategy requires maximizing the objective function given by Theorem~\ref{thm:success}. Unfortunately, this objective function is complex, and this makes the voters' general optimization problem in \eqref{eq:max_prob_alg} challenging. 
To understand the origin of this complexity, we visualize in Figure~\ref{fig:threshold_few_voters} the objective function, that is, the probability of electing an honest committee.

Suppose for now that \emph{all} voters follow a threshold voting strategy, with some common threshold, $z$.  In this setting, we can calculate the \emph{exact} probability of success as a function of the threshold chosen. Figure~\ref{fig:threshold_few_voters} shows the success probability under threshold voting, for \emph{small} numbers of voters ($n=1$ to $n=4$).  Although, in practice, systems have many more voters, these graphs highlight some of the complex dynamics of approval voting.

The optimal thresholds tend to hover around $0.5-0.7$, meaning that with these parameters, voters should vote for any candidate, $j$, whose posterior probability, $s_{ij}$, is above this threshold and not vote for any candidate below this threshold. The thinness of the peaks, however, indicates that even small deviations from the optimal strategy can drastically reduce the success probability. In addition, the number of local optima appears to increase with $n$. These properties make the general optimization problem intractable at relatively low or medium values of $n$ (with the exception of $n=1$, for which the objective is unimodal).

\begin{figure}[H]
    \centering
    \includegraphics[width=.5\textwidth]{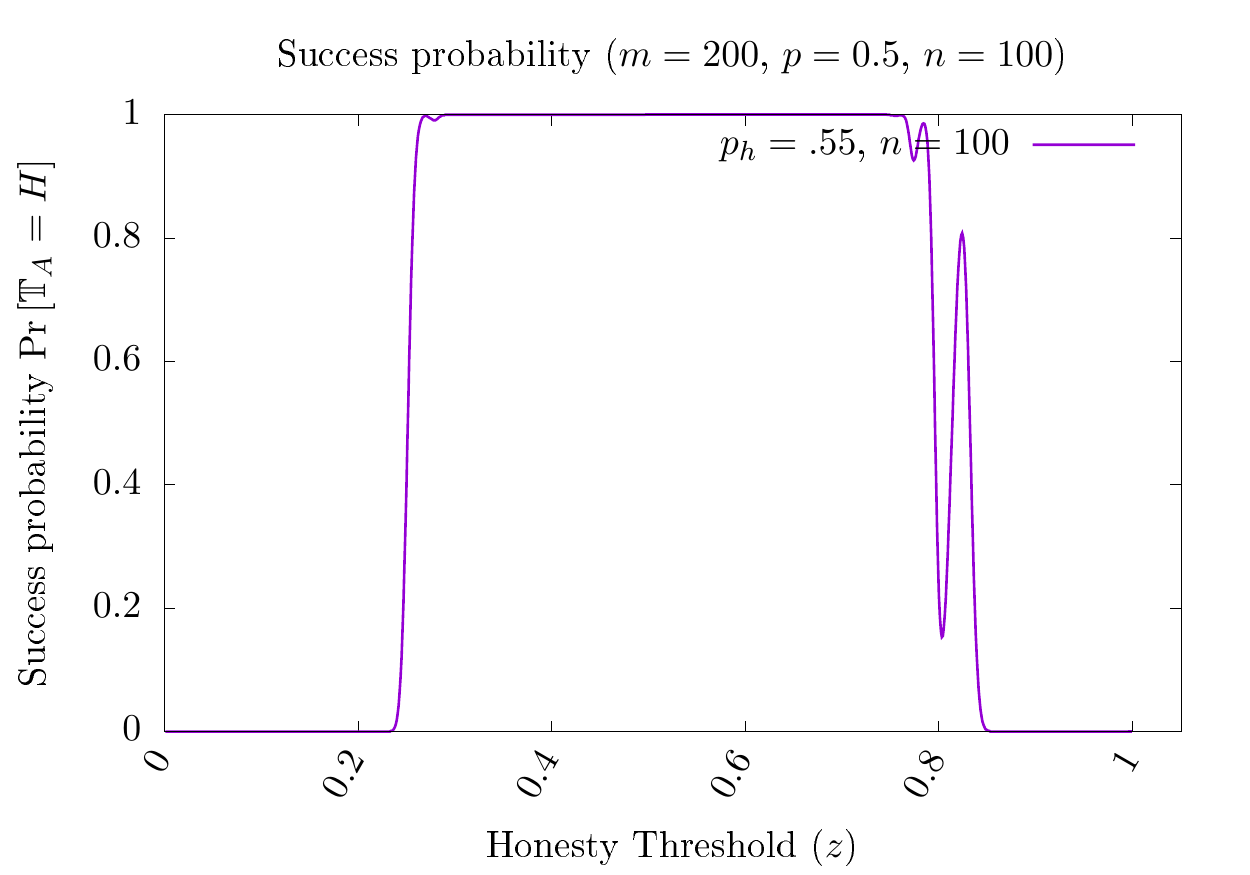}
    \caption{
    Success probability as a function of threshold chosen, assuming large number of voters ($n=100$). \emph{Key Takeway:} For large $n$, the probability of success goes to 100\% across a wide range of thresholds, and thus a wide range of voting strategies yields nearly optimal results.
    \label{fig:threshold_100_voters}}
\end{figure}

Next, we examine the situation for a large number of voters $n=100$, in Figure~\ref{fig:threshold_100_voters}.

Figure~\ref{fig:threshold_100_voters} shows that for large $n$, the previous issues fade: the local optima tend to merge, and almost any reasonable threshold has a close to 100\% probability of success. 

Combining insights from Figures~\ref{fig:threshold_few_voters}  and ~\ref{fig:threshold_100_voters}, we conclude that the voters' problem behaves drastically differently depending on low vs. high-number of voters, and hints that asymptotic analysis may offer more tractable results.

Next, we analytically characterize (to the extent possible) the optimality of voting strategies, separating the $n=1$ case from the general $n>1$ case.

\subsection{Special Case: Equivalence between $n=1$ \& $n>1$ with Signal Pooling}
\label{sec:single_voter}

In this section we consider the special case of a single voter ($n=1$). Beyond letting us build intuition, we show that the general $n>1$ case, collapses to $n=1$ when voters are allowed to credibly (and costlessly) share their signals. This is more than a mere hypothetical exercise. Signal pooling has no obvious downside in our setting given voter incentives are aligned, and thus could be plausible in practice.

\subsubsection*{Single Voter}
Before presenting the result, we introduce one intermediate technical lemma that will be useful
throughout the analysis.

\begin{lemma}
    \label{lem:poisson_maximality}
    Suppose $X$ is a Poisson Binomial random variable with $k$ trials and let $X_P$ with $P=\{p_{j_1},\ldots,p_{j_k}\}$ denote the Poisson Binomial with parameters $(p_{j_1},\ldots,p_{j_k})$. Let $m\geq k$, and $x$ be positive integers, $\{p_1,\ldots,p_m\}\subseteq [0,1]^m$ such that $p_1\geq \cdots\geq p_m$ then 
    $\underset{P\subseteq \{p_1,\ldots,p_m\}}{\operatorname{argmax}} {\Pr [X_P>x]}=\{p_1,\ldots,p_k\}$.
\end{lemma}

To understand the implication of Lemma~\ref{lem:poisson_maximality}, consider a $k$-winner approval voting system in which candidates have posterior probabilities of being honest $s_1,\ldots,s_m$. Suppose the subset of candidates elected to the committee is $\mathbb{T}=\{c_{j_1},\ldots,c_{j_k}\}$ so that their posterior probabilities are $\{s_{j_1},\ldots,s_{j_k}\}$. If we think about the realization of honesty of each candidate as a trial $l$ with probability $s_{j_l}$ then the number of honest candidates on the committee $X$ is a Poisson Binomial with parameters $(s_{j_1},\ldots,s_{j_k})$. The success probability, e.g. the probability of an honest committee is the probability that the number of honest candidates on the committee is at least $\inceil{(1-\bp)\cdot k}$. Lemma ~\ref{lem:poisson_maximality} implies that the success probability is maximized when each of the posterior probabilities of the different candidates is as high as it can be.

\begin{figure}[H]
    \centering
    \includegraphics{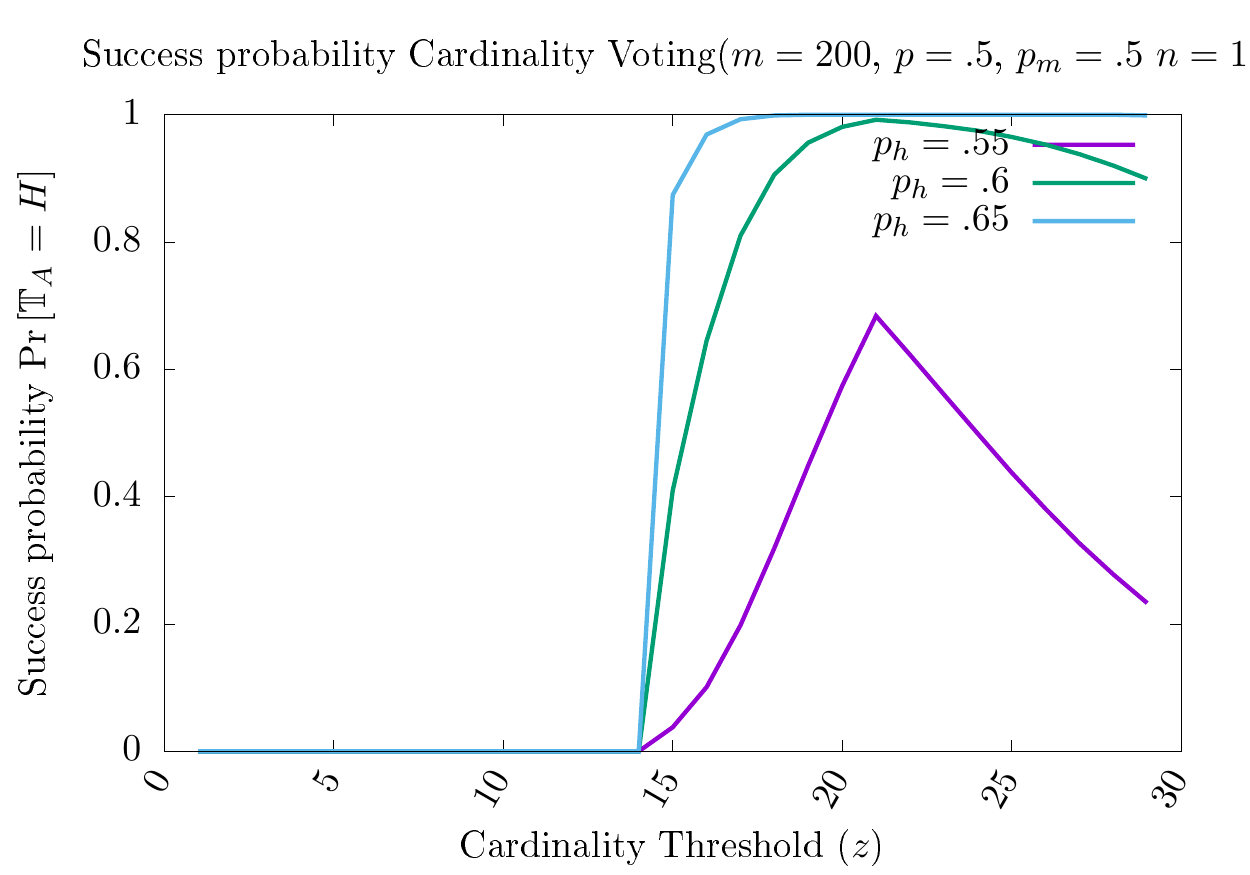}
    \caption{The success probability when there is a single voter who follows the cardinal voting strategy.  In this figure, the committee size, $k = 21$.  The optimal success probability occurs when the threshold $z = k$, even as the accuracy of the voter changes.}
    \label{fig:single_voter_cardinal}
\end{figure}

\begin{proposition}[Optimality of Cardinal voting when $n=1$]
	\label{prop:single}
	Consider a $k$-winner approval voting system with $n=1$ voter and $m\geq k$ candidates, then the globally optimal strategy is the cardinal strategy with $z=k$.
\end{proposition}

Proposition~\ref{prop:single} follows from the fact that if there is only a single voter, that voter possesses \emph{all} relevant information about each producer's type, and the voter can unilaterally decide the committee.  Thus the optimal strategy is to form the committee from the candidates that have the highest (posterior) probability of being honest.  In other words, the voter should vote for the top $k$ candidates (when sorted accorded to their posterior probability of being honest), and this strategy is optimal across \emph{all} possible strategies, not just cardinal or threshold voting.

\begin{proposition}[Suboptimality of Threshold voting when $n=1$]
	\label{prop:threshold_suboptimal}
	Consider a $k$-winner approval voting system with $n=1$ voter and $m\geq k$ candidates, then any threshold strategy is {\bf not} optimal.
\end{proposition}
More specifically, when we say a strategy $A\in \mathbb{S}$ is not optimal we mean that there is a non-zero probability event (realization of signals) in which the non-optimal strategy achieves a success probability that is strictly smaller than would be achieved using a different strategy $B\in \mathbb{S}$. 
\begin{equation}
    \label{eq:non_optimal}
 \Pr \inbrak{\mathbb{T}_{A}=H\suchthat (\mathbf{s}_i)_{i=1}^n}<\Pr\inbrak{\mathbb{T}_{B}=H\suchthat (\mathbf{s}_i)_{i=1}^n}\text{, for }\Pr[(\mathbf{s}_i)_{i=1}^n]>0.
\end{equation}
In particular, in the proof of Proposition~\ref{prop:threshold_suboptimal} we show Equation (\ref{eq:non_optimal}) is true for $n=1$ and $A=\text{Threshold},B=\text{Cardinal}$. In other words, we show that for $n=1$ the threshold strategy gives a strictly lower success probability than the cardinal strategy.

\subsubsection*{Signal Pooling}\label{sec:sigpool}

If voters could credibly (and costlessly) share their private signals, then it is straightforward to show that they effectively act as a single voter.

As in the private signal setting, each voter can calculate the probability that a given producer is honest, conditioned on the received signals.  But now, we assume voters can condition on \emph{all} the signals.  We calculate the resulting posterior probability in Lemma~\ref{lem:posterior_multiple} in Appendix~\ref{app:posterior}.

Proposition~\ref{prop:multiple} shows that when voters share their signal, the optimal strategy (out of \emph{all} possible strategies) is to follow the cardinal voting strategy with threshold $z = k$.

\begin{proposition}[Optimality of Cardinal Voting with Shared Signals]
    \label{prop:multiple}
    Consider a $k$-winner approval voting system with $n>0$ voters, $m\geq k$ candidates and such that voters' private signals are credibly shared. Then the globally optimal strategy is the cardinal voting strategy with $z=k$ where each voter $v_i$ is ranking based on the shared $s_j$ instead of their private $s_{ij}$.
\end{proposition}

A natural question that follows is, whether cardinal voting persists to be the optimal strategy in the general multi-voter case. We examine this in the next section.

\subsection{General Case: Multiple Voters}
\label{sec:multiple_voters}

Given we show in the previous Section~\ref{sec:single_voter} that the threshold strategy is already suboptimal for $n=1$, while the cardinal strategy is in fact optimal for $n=1$, we focus our attention here on the latter.

In the multi-voter setting ($n>1$), the optimal cardinal strategy becomes extremely complex and even computing the exact success probability for a fixed strategy is difficult. Despite this, we can formally show that the cardinal strategy that was always optimal with $n=1$ may become suboptimal with $n>1$. 

\begin{proposition}[Suboptimality of Cardinal voting when $n > 1$]
\label{prop:cardinal_suboptimal}
Consider a $k$-winner Approval Voting system with $n>1$ voter and $m\geq k$ candidates, then the cardinal strategy can be suboptimal.
\end{proposition}

Intuitively, this result occurs because a vote for a candidate can actually \emph{bump} other candidates out of the committee. To dig deeper, consider a situation with two voters (voter $0$ and voter $1$), where voter $0$ has better information than voter $1$,
(i.e., $\sigma_{0j} \ll \sigma_{1j}$ for all $j \in [m]$).
Even if $\sigma_{1j}$ is large, voter $1$'s signals convey information (a single voter who had access to the signals $(s_{01},\ldots,s_{0m})$ \emph{and} $(s_{11},\ldots,s_{1m})$ 
would do better than one with access to $(s_{01},\ldots,s_{0m})$ alone).
The problem is that voter 1 can only convey information about their signal through \emph{discrete} votes, 
and a vote for candidate $j$ may be too strong an endorsement for that candidate given that the signal $s_{1j}$ is only weakly informative.

As an extreme case, consider a situation where voter $0$ is perfectly informed (i.e., voter $0$ can differentiate between honest and dishonest producers with probability 1), 
and voter $1$ is perfectly uninformed (i.e., from voter $1$'s perspective, each candidate is honest independently with probability $p$, in other words $s_{1j} = p$ for all $j \in [m]$ ).
In this case, it should be clear that voter $1$ should \emph{not} cast any votes, while voter $0$ should cast $k$ votes.

The suboptimality of the cardinal voting strategy persists even if both voters have the same information (i.e., $\eps_{0j} = \eps_{1j}$ for $j \in [m]$).
This is because the \emph{realized} signals can convey different amounts of information.  
For example, suppose voter $0$'s sorted signals are $s_{0j^0_1} \ge s_{0j^0_2} \ge \cdots \ge s_{0j^0_m}$ 
and voter $1$'s sorted signals are $s_{1j_1^1} \ge \cdots \ge s_{1j^1_m}$.
Suppose as well that $s_{0j^0_k} \gg s_{0j^0_{k+1}}$, but $s_{1j^1_{k-1}} \approx s_{1j^1_{k}} \approx s_{1j^1_{k+1}}$.
In this case, voter $0$ has high confidence that the committee should consist of the candidates $\inset{ j^0_1,\ldots,j^0_k }$, 
but voter $1$ is essentially indifferent between candidates $j^1_{k-1},j^1_k,j^1_{k+1}$.
The cardinality strategy with threshold $k$ would force voter 1 to vote for $j^1_{k-1}$ and $j^1_k$, but not $j^1_{k+1}$, 
and this vote (based on little information) could displace committee members who would have been elected by voter $0$ (whose signals were very informative).

\subsection{Asymptotic Optimality (Main Result)}\label{sec:asymptotic}

So far, we have established via Proposition~\ref{prop:threshold_suboptimal}, that the threshold strategy is suboptimal, and via Proposition~\ref{prop:cardinal_suboptimal}, that 
the cardinality strategy may be suboptimal. However, our basic numerical study in Section~\ref{sec:complexity} suggests that as the number of voters increases, the optimality gap may decrease.

\begin{theorem}[Exponential Convergence (Main Result)]
\label{thm:approximate_optimality}
	Let $\cM$ denote the set of dishonest producers, 
	and suppose there exists a set of honest producers, $\cH$, with $|\cH| \ge (1-\bp) \cdot k$,
	and a $\delta > 0$ such that
	$
		\min_{i \in [n], j \in \cH} p_{ij} \ge \max_{i \in [n], j \in \cM} p_{ij} + \delta,
	$
	where $p_{ij}$ denotes the probability that voter $i$ votes for producer $j$.
	Then the probability of electing an honest committee is bounded from below and tends to 1 as $n$ increases, in particular, we have
	\begin{equation}
		\Pr \inbrak{ \mathbb{T} = H} \ge 1 - 2m^2 e^{-\delta^2 n/2}  \underset{n \rightarrow \infty}{\quad \xrightarrow{\hspace*{0.6cm}} \quad  1.}
	\end{equation}
\end{theorem}

Theorem~\ref{thm:approximate_optimality} shows that as the number of voters, $n$, tends to infinity, then the probability of electing an honest committee tends to 1, exponentially quickly, as long as the signals are not completely uninformative, that is, as long as there exists an arbitrarily small $\delta>0$ gap s.t. $p_{ij} > p_{ij'} + \delta$ when $j$ is honest and $j'$ is dishonest.  The lower bound on the success probability \emph{decreases} quadratically in the number of block producer candidates $m$, because if there are too many block producers relative to the number of voters, no single block producer can amass enough votes to make it onto the committee with high probability.  As long as there are not too many candidates, however, the exponential dependence on the number of voters dominates the quadratic dependence on the number of producer candidates. Importantly, this result holds across all voting strategy classes, and this leads to the following two corollaries.

\begin{corollary}
    \label{cor:threshold_convergance}
	If $p_h > p_m$, and $m \cdot p \gg k$, then the probability of an honest committee 
	when all voters follow the Threshold strategy converges to $1$ as $m, n \rightarrow \infty$
	(assuming $n \gg m$).
\end{corollary}

\begin{corollary}
    \label{cor:cardinal_convergance}
	If $p_h > p_m$, and $m \cdot p \gg k$, then the probability of an honest committee 
	when all voters follow the Cardinality strategy converges to $1$ as $m, n \rightarrow \infty$
	(assuming $n \gg m$).
\end{corollary}

\begin{remark}
    Since $m$ is the number of producers, and $p$ is the \textit{a priori} probability a candidate is honest, 
    the number of honest candidates is distributed as $\Bin(p,m)$, and the expected number of honest candidates is $m \cdot p$.  If the number of honest candidates is less than $k$, then there is no way to elect $k$ honest producers.  The assumption that $m \cdot p \gg k$ ensures that the probability there are fewer than $k$ honest candidates is small.
\end{remark}

\begin{figure}[H]
    \centering
    \includegraphics[width=.5\textwidth]{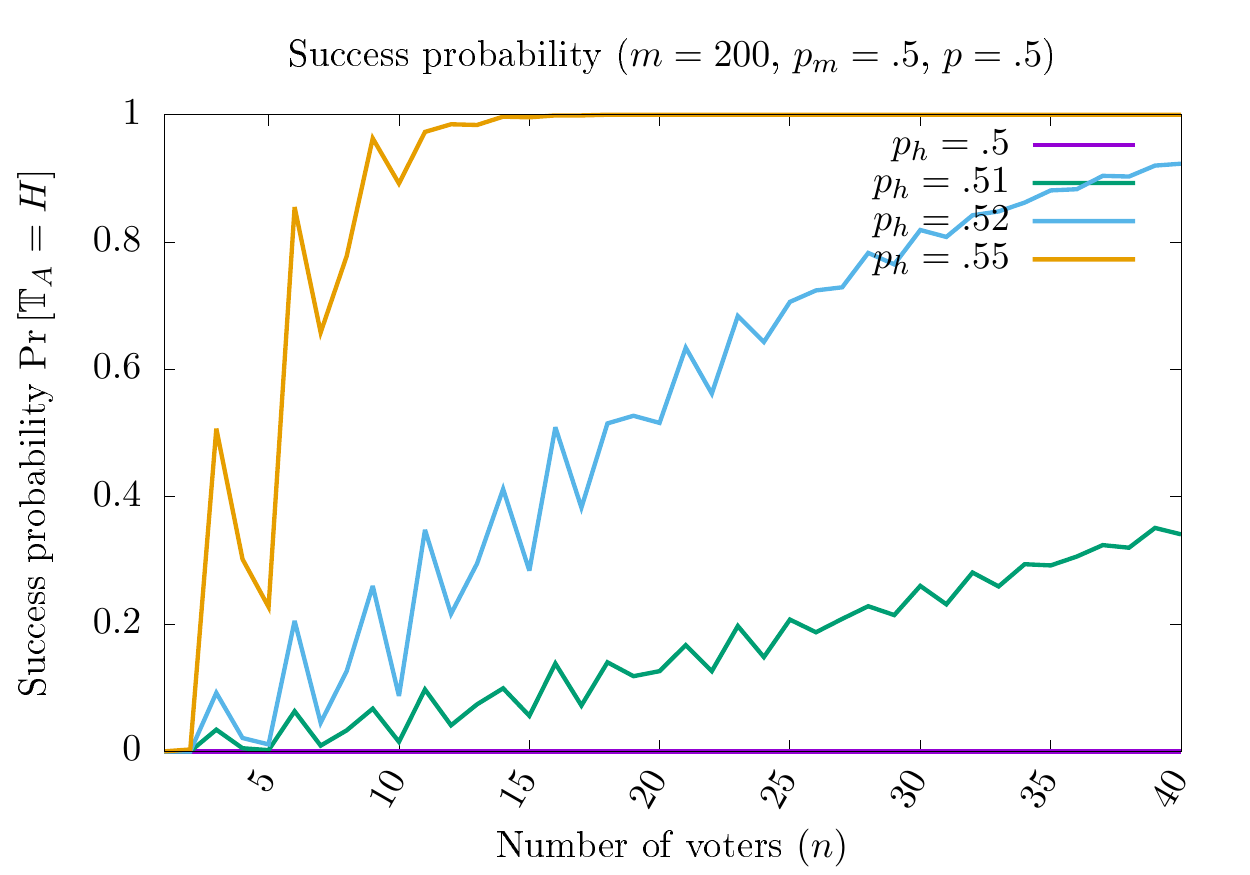}
    \caption{Rate of Convergence to Optimality under a suboptimal threshold voting strategy. Note that the figure shows \emph{exact} probabilities---the line jaggedness comes about from the combinatorial nature of the voting problem.  {\bf Key Takeaway:} The success probability quickly goes to one, as voters $n$ increase.}
    \label{fig:order_stats_n}
\end{figure}

Figure~\ref{fig:order_stats_n} illustrates the exponential convergence result, assuming each voter follows a (generally suboptimal) Threshold Voting strategy (Definition~\ref{def:threshold}) with $z = p$. The figure shows that even with relatively weak signals, the probability of success rapidly converges to 100\%.

Note, convergence to optimality also holds (more trivially) for other asymptotics of interest, such as if the signal informativeness $\frac{p_h-p_m}{\sigma} \rightarrow \infty$ or if the prior $p \rightarrow 1$. See Appendix~\ref{app:other_asymptotics} for an illustration.

\section{Efficiency vs .Security: Comparison of Consensus Protocols}
\label{sec:PoScomparison}

In this section, we compare the approval voting mechanism analyzed in \S\ref{sec:results}, to two other popular mechanisms implemented in practice: single-choice voting (e.g., Cosmos and Binance Smart Chain), and lottery based (e.g., Algorand).

\subsection{Single-Choice Voting (e.g., Cosmos, Binance, etc.)}

Most blockchains built on the Cosmos SDK employ single-choice committee-based election,
and the elected committee runs the Tendermint consensus protocol (which requires a $1/3$ fraction of honest participants).  This is the voting mechanism used by the Cosmos Hub, crypto.org, Osmosis, Secret, Oasis, Binance Chain etc.  It is also the mechanism used by Tron and the Binance Smart Chain.

Mathematically, the single-choice setting is equivalent to our cardinal voting setting \emph{where voters would be required to set their cardinality threshold $z = 1$}. In this case, each voter's optimal strategy is intuitive: vote for the candidate that is most likely to be honest, \ie voter $i$ votes for candidate $j$ where $s_{ij}$ is maximized. Despite this, one cannot directly apply Theorem~\ref{thm:success} to recover the closed-form aggregate success probability, because the number of votes received by each candidate are \emph{not} independent variables (if voter $i$ votes for candidate $j$, voter $i$ does \emph{not} vote for candidate $j'$).  We therefore analyze this protocol numerically focusing on the following question: how does the $z=1$ restriction affect the protocol's success probability (compared to setting a different global threshold $z >1$)?

\begin{figure}
    \begin{center}
        \includegraphics[width=.8\textwidth]{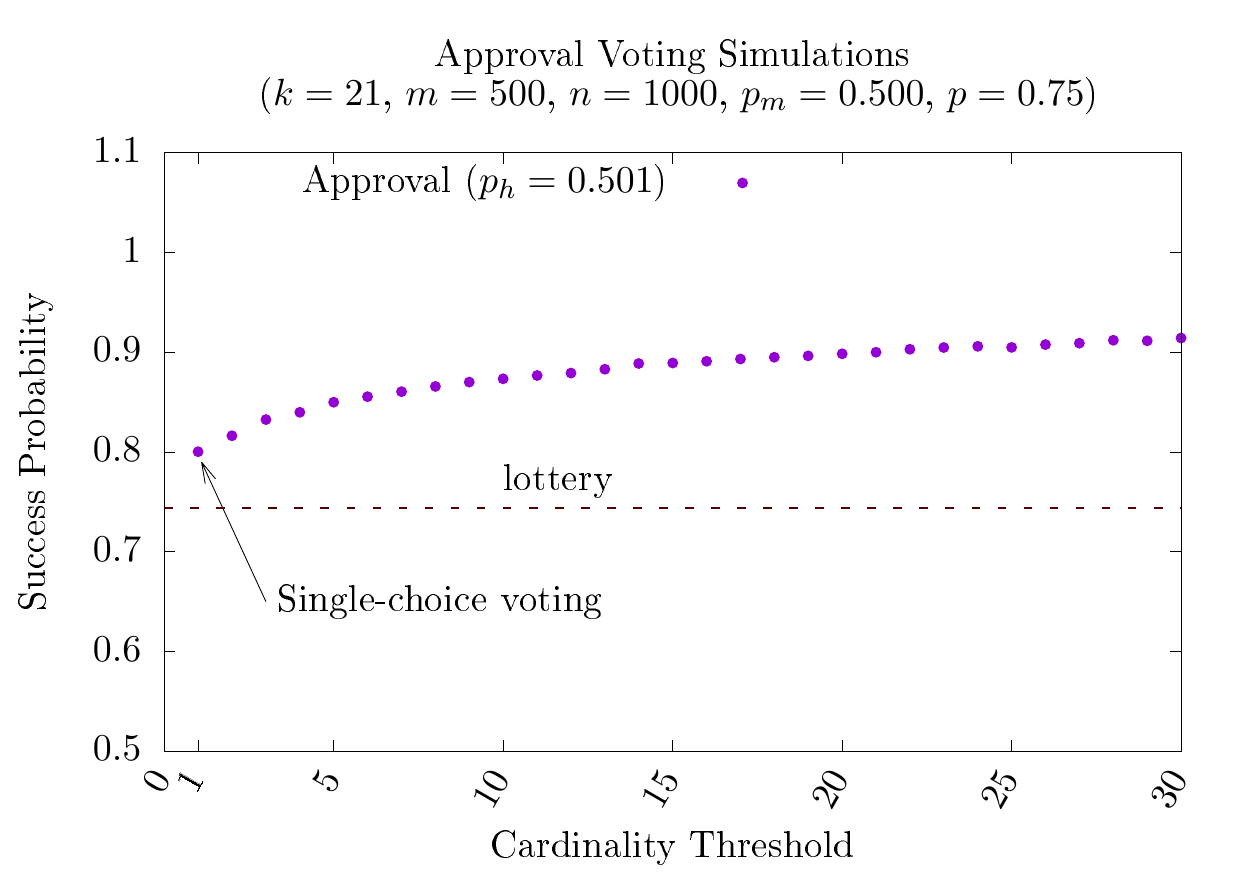}
        \caption{Success probabilities with cardinal voting strategies.  When $z=1$ (i.e., single-choice voting) the success probability is at its lowest.  {\bf Key takeaway:} Voting for a single candidate (which is the only possible strategy on most platforms) is essentially the \emph{worst} strategy. \label{fig:cardinalsims21}}
    \end{center}
\end{figure}

Figure~\ref{fig:cardinalsims21} shows how the probability of electing an honest committee differs when voters follow a cardinal voting strategies with different thresholds, $z$.  The key observation is that when voters vote for only a single voter, $z = 1$, the success probability is minimized, that is, the single-choice restriction constitutes the \emph{worst} possible cardinal voting strategy.

By contrast, in more general approval voting systems (e.g. EOS, Telos), even if voters choose their threshold suboptimally, they are essentially guaranteed to have higher success probabilities than in the single-choice setting.

The lottery protocol which bypasses voting in favor of random selection (e.g., Ethereum) is represented by the dashed line. We assume $p = .75$ (i.e., each candidate has a 75\% chance of being honest) and the committee size is 21.

\subsection{Lottery-based committees (e.g., Algorand)}
\label{sec:lottery}

Algorand employs committee-based consensus, but uses a randomly selected committee to certify each block.
In order to ensure that the Algorand protocol never forks, the protocol must \emph{never} elect a dishonest committee.

The analysis of the Algorand protocol proceeds as follows.
Suppose some fraction, $p > 2/3$, of the tokens are held by honest participants 
(\ie, an elected committee member is honest with probability $p$) then if a committee of size $k$ is randomly selected, then the 
probability the committee is at least $2/3$rds honest is 

\begin{equation}
    \Pr \inbrak{ \Bin{k,p} \ge \frac{2}{3} k. }
\end{equation}

A Chernoff bound (e.g. \cite{DP09}[Exercise 1.2]) then shows

\begin{equation}
    \Pr \inbrak{ \Bin{k,p} \ge \frac{2}{3} k} > 1 - e^{-\frac{ \inparen{ 1-  \frac{2}{3p} }^2 p k }{2}. }
\end{equation}

which decays exponentially as the committee size, $k$, increases (as long as $p > \frac{2}{3}$).  In particular, we can make the failure probability arbitrarily small by choosing the committee size to be large enough. 

\begin{remark}
    A key drawback of the lottery-based election method is that the success probability only converges to 1 \emph{as the committee size ($k$) increases}.  By contrast, Theorem~\ref{thm:approximate_optimality} shows that for approval voting, the success probability converges to 1 \emph{as the number of voters increases}.  This is the key reason why voting-based systems can have much smaller committee sizes than lottery-based systems.
\end{remark}

For example, Algorand suggests a target committee size of about $1500$, instead of 21 for EOS \citep{algorand}[Section 5.1].
If we assume that (at most) $20\%$ of the tokens are ever held by malicious participants, a Chernoff bound gives that the 
probability of electing a dishonest committee (\ie a committee with more than $1/3$ dishonest members), is bounded by 
$10^{-12}$.  

The probability that a dishonest committee is \emph{ever} elected can then be bounded by taking a union bound 
over all potential elections (e.g. if there is an election every four seconds for the next twenty five years, there will be 
approximately $200$ million elections).
Taking a union bound over the $2 \cdot 10^8$ elections held in the text 25 years, we have the probability 
of a fork in the next 25 years is at most $.5\%$.

In Algorand, it is very easy to calculate the probability of a dishonest committee, for a given fraction of honest candidates ($p$), and a given committee size ($k$). Intuitively speaking, allowing users to vote should increase the probability of electing an honest committee, and thus reduce the size of the committee needed to ensure that it reaches the critical ($2/3$rd) threshold of honest members.  

\begin{figure}[H]
    \centering
    \includegraphics[width=.5\textwidth]{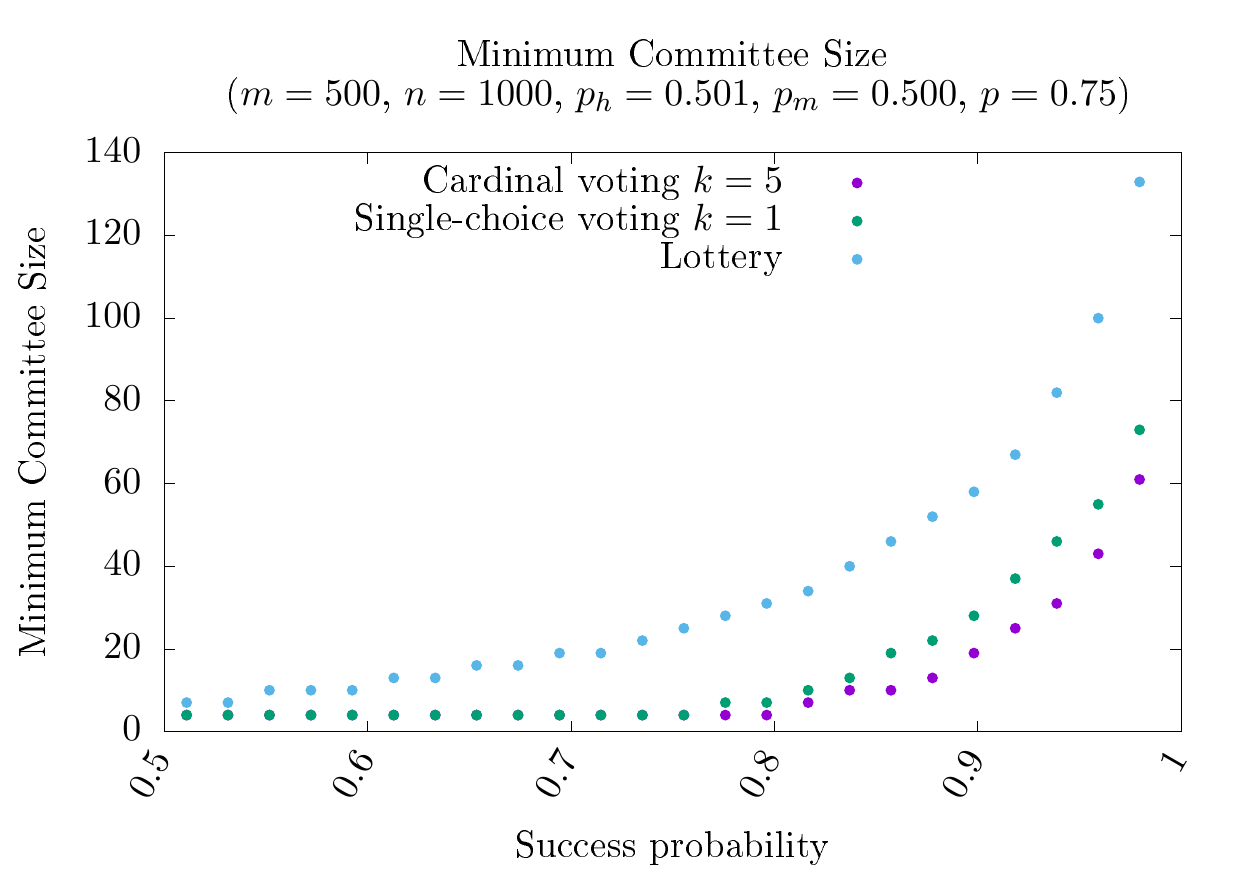}
    \caption{The minimum committee size required (y axis) to achieve a given success probability, when the committee is chosen by lottery (as in Algorand) vs. when the committee is elected by single-choice voting (e.g. Cosmos) or general Cardinality voting with $k=5$. Each point in the cardinal voting strategies represents the mean success probability from 100,000 samples.  The committee-sizes for lottery-based consensus can be solved analytically. {\bf Key takeaway:} electing committees by on-chain voting requires much smaller committee sizes for the same level of security.}
    \label{fig:committee_size}
\end{figure}

In Figure~\ref{fig:committee_size}, we plot the minimum committee size necessary to achieve a desired failure probability, when the committee is chosen by lottery (as in Algorand or Espresso) or according to an approval vote (as in DPoS).  
The key takeaway is that even when the voters have only minimal information $(p_m = .5, p_h = .501$ and $\sigma = .1$), allowing users to vote for candidates \emph{drastically} reduces the size of the committee necessary to achieve a specific failure bound.  Since the committee executes a Byzantine Agreement protocol with communication cost that is quadratic in the committee size, $k$, minimizing the committee size is critical for performance.

Note, Figure~\ref{fig:committee_size} was generated assuming the voters follow a \emph{cardinal} voting strategy.  When $z = 1$, this is single-choice voting, i.e., the case when each voter is only allowed to vote for one candidate (as in most Cosmos-based chains).  When $k = 5$, this is the case when voters vote for their top 5 candidates.  This is \emph{not} the optimal cardinal voting strategy, but this plot shows that even when voters follow a simple, but suboptimal voting strategy, election-based committee selection can yield very small committees and outperform the other protocols. To summarize:

\begin{result}
Approval voting outperforms both lottery-based, and single-choice voting consensus mechanisms, even when users are assumed to follow suboptimal cardinality thresholds.
\end{result}

\section{Discussion and Conclusion}\label{sec:fairness}

The paper analyzes committee-based consensus protocols and shows that even though elections based on approval voting lead to intractable optimal voting strategies, they nonetheless exhibit strong robustness to more intuitive (suboptimal) voting strategies that users resort to in practice, converging to optimality exponentially quickly in the number of voters. The paper also shows that committee-based consensus using approval voting has the potential to outperform other commonly employed mechanisms such as single-choice and lottery-based protocols (in terms of failure rates). We next discuss some of the limitations.

\textbf{Stake-weighted voting:}
    In our model, all voters votes have equal weight, whereas many platforms rely on stake-weighted voting. In this case, the Poisson Binomial Distributions become Generalized Poisson Binomials \cite{GeneralizedPBD}.
	In particular, Theorem~\ref{thm:success} holds as is, but the distribution functions (defined in Theorem~\ref{thm:voting_distribution}) become somewhat more involved.
	Importantly, the qualitative insights of the paper (e.g., asymptotic optimality) continue to hold.

\textbf{Different objectives:}
 Beyond optimizing to reduce failure rates, our model does not deal with other features that voters may care about. These are outside the scope of this work, but could be of interest for future work.  

For instance, one drawback of electing committees as opposed to selecting random committees is that elections seems to lead to stagnation, especially early on in the blockchain life-cycle.  EOS represents a rather extreme example: The first 89 million EOS blocks were mined by only 63 distinct producers \cite{Xblock}.  By comparison, the first 655,000 Bitcoin blocks were mined by more than 275,000 distinct addresses, and the first 8 million Ethereum blocks were mined by over 5000 distinct addresses \cite{xblock-eth}. 

A small, static set of block producers reduces decentralization -- a core tenet of almost all cryptocurrencies. The idea that there should be a diversity of block producers is core to the open, democratic ideals that spawned much of the blockchain ecosystem, and the idea that there should be turnover in the set of block producers has been formalized in the notion of  \emph{chain quality} which is a measure of fairness.  Chain quality is a measure of whether (in sufficiently long time windows) the fraction of blocks contributed by each participant is proportional to their hash power or stake \cite{GKL15}.

Chain quality is a different metric by which we could measure different election mechanisms, and this could be an interesting direction for future research. A discussion can be found in Appendix~\ref{sec:quality}.

\textbf{Alternative voting schemes:}
In this work, we focused on lottery-based selection, single-vote mechanisms and approval voting because these are the systems that have currently been deployed for committee selection.

Other alternative voting systems exist, e.g. ranked-choice voting \cite{rankedchoice2021}. Even though we were not able to find practical implementations, it could be interesting to analyze whether these can outperform approval voting in settings where voter incentives are aligned, but voters are (on the whole) poorly informed.

A separate question is vote weighting.  All current Proof-of-Stake blockchains weight stake \emph{linearly}, but there are alternative weighting mechanisms, the most common being \emph{Quadratic Voting} \cite{PW14,LW18}.  In Quadratic Voting, a voter's vote weight is proportional the \emph{square-root} of their stake, rather than being proportional to the stake itself.  Quadratic Voting has been used in the blockchain context (e.g. Gitcoin Grants), but to the best of our knowledge, has \emph{not} been used as a method for electing a consensus committee.

\bibliographystyle{informs2014}
\bibliography{main.bib}

\newpage
\appendix
{\centering \Huge Appendix}

\section{A Primer on Blockchain Consensus Protocols \& Committee Elections}\label{sec:primer}

Herein we provide a primer on single-leader and committee-based consensus protocols. Readers familiar with their basic operational features can skip this section without loss.

As mentioned in the introduction, the core problem facing all cryptocurrencies (and decentralized databases of all kinds), is how to provide a single, universally accepted \emph{ordering} of transactions (or state updates). Most modern cryptocurrencies are based on the notion of a \emph{hash chain}, where blocks of data are chained together using cryptographic hash functions.  Hash chains are an append-only data structure, meaning that new blocks (containing transactions) can be appended to the end of the chain, while internal blocks of the chain cannot be modified or re-ordered (without modifying all subsequent blocks). Since anyone can easily append new blocks to the end of a hash chain, decentralized systems need a method for deciding how and when new blocks can be added to the chain.

Most cryptocurrencies use a form of leader election, where a leader is elected at regular intervals. This leader, or ``block producer,'' is given the right to produce a single block. There is an inherent value in becoming a block producer, as block producers have the power to insert, re-order and censor transactions \citep{flashboys2.0}.\footnote{The value that can be extracted by inserting, re-ordering and censoring transactions is termed  ``Maximum Extractable Value" (MEV), and is worth hundreds of millions of dollars on blockchains like Ethereum \citep{MEVexplorer}.}
In addition, most cryptocurrencies provide direct incentives for block production  in the form of transaction fees and block rewards. Transaction fees are paid by the users to incentivize the block producer to include specific transactions in a block.  Block rewards are new coins that are minted and paid directly to block producers.  For example, in Bitcoin, block rewards are currently set to $6.25$ BTC. In many cryptocurrencies, block rewards are the \emph{only} mechanism by which new coins are generated. 
For reference, in July 2021, Ethereum miners received about 18\% of their direct compensation from transaction fees and 82\% from block rewards \citep{Theblock}.

Block producers also have the ability to harm the platform itself.  Block producers can censor transactions \emph{within the block they produce}.
Lazy or inept block producers can reduce the total transaction throughput of the system by failing to include enough transactions in a block or failing to produce a block altogether. Malicious block producers can ``fork'' the chain by appending two blocks at the same block height.  This type of behavior can lead to ``double-spending attacks'' and can destabilize the entire blockchain.

Block producers' power to harm the ecosystem, means that the selection mechanism  must ensure that only ``honest'' producers are elected. When block producer candidates have stable identities, classical consensus protocols (e.g. \cite{LSP82,CL99}) provide efficient and robust mechanisms for leader election. In a permissionless setting, however, where the set of block producer candidates is anonymous and dynamic, classical consensus protocols fail, and other methods must be devised.

\subsubsection*{Proof of Work (PoW)}

As mentioned earlier, the Bitcoin whitepaper \cite{nakamoto2008bitcoin} introduced a novel single-leader protocol whereby block-producing candidates (``miners'') expend effort in the form of computing cryptographic hashes on random values, and their chance of becoming block leader is proportional to the amount of effort they exert.  This Proof of Work consensus, is used by many of the leading cryptocurrencies (by market cap), including Bitcoin, Ethereum, Dogecoin and Litecoin. 

Although PoW-based consensus has proven stable and secure, it has several drawbacks, most notably its societal cost, and its low transaction throughput. Currently, block producer candidates on Bitcoin expend about as much electricity as the country of Finland in an effort to be chosen as block producers \citep{cambridge}.

Proof-of-Work based consensus, also has limitations on how frequently block producers can be chosen, and this directly affects the blockchain's transaction throughput. Currently, the leading PoW-based blockchains, Bitcoin and Ethereum, can handle less than tens of transactions per second. By contrast Visa handles thousands of transactions per second \citep{tps}.

\subsubsection*{Proof of Stake (PoS)}

The aforementioned drawbacks of PoW have pushed the blockchain community to explore alternatives, and in this quest, Proof of Stake (PoS) has arguably emerged as the current frontrunner. The Ethereum blockchain, for instance, which supports the world's second largest cryptocurrency ETH, originally launched with a PoW protocol but has been gradually trying to transition to a form of PoS for several years \citep{ethmerge}.

In PoS, block producers are elected in proportion to their token balance (``stake'') on the blockchain, rather than their computational effort.  Similar to PoW systems, where candidates signal their support of the platform by expending computing resources, in PoS systems, candidates signal their support of the system by acquiring and holding native tokens on the blockchain.

There are many variants of the PoS protocol, but a common feature of almost all PoS systems is that block producers are elected with probability proportional to their ``staked'' tokens (as in Ethereum 2.0) or their passive token balances (as in Algorand).

\subsubsection*{Committee-based Consensus}

Although under PoS, block producers can earn significant returns, being an efficient block producer usually requires powerful computing equipment, a dedicated internet connection, and a robust software configuration. In some blockchains, a nontrivial minimum amount of tokens is also required to be eligible to participate. Many regular token holders are thus ineligible (or simply unwilling) to take on this type of role. To address this problem, most PoS systems support some type of delegation mechanism, whereby token holders can delegate their stake to professional block producers (usually in exchange for some sort of profit sharing).

Committee-based consensus takes this separation between token holders and block producers to the extreme. In most traditional PoW systems, block producers are selected in a lottery-like procedure according to their (proportional) hash power.  Several PoS systems (e.g. Tezos, Algorand, Cardano) adapted this idea to elect leaders randomly with probability equal to their proportional token stake.\footnote{%
In fact, the core technical contribution in systems like Algorand and Cardano is a decentralized, verifiable lottery mechanism.} As an alternative to this lottery-based leader election, several blockchains allow users to cast (stake-weighted) votes for block producers and the ones with the highest number of votes become producers for some fixed duration of time.

In platforms using committee-based consensus, a small committee ($k = 150$ in the Cosmos Hub, $k = 21$ in EOS or $27$ in TRON) of block producers is elected by a stake-weighted vote, and is responsible for producing and validating blocks.\footnote{Although almost all PoS systems support some form of delegation, the term ``Delegated Proof of Stake'' is usually reserved for the specific type of committee-based consensus protocols used by systems like EOS and TRON.}

In committee-based consensus, the elected committee typically runs a traditional consensus algorithm --- Practical Byzantine Fault Tolerance \citep{CL99}, Proof-of-Authority \citep{PoA} or Tendermint \citep{Tendermint} --- to certify the next block.   

\subsubsection*{Some Advantages of Committee-based Consensus}
Committee-based consensus has several perceived advantages over other commonly used consensus protocols.  
First, the committee can check each other's actions and prevent malicious behavior.  For example, in most classical consensus protocols up to a $\bp$-fraction\footnote{Most consensus protocols can tolerate $\bp = 1/3$.} of the participants can behave maliciously without adversely affecting the system.  

Second, the voting process takes a nonzero amount of time, so electing a batch of producers at once increases efficiency in contrast two Nakamoto consensus, where a single block producer is selected at each step.

Third, it allows the chain to achieve instant finality -- when the committee certifies a block, that block is immediately finalized.  This is in contrast to PoW blockchains that only achieve eventual finality. Bitcoin wallets, for instance, typically wait until a transaction is buried 6 blocks deep in the chain before considering it ``finalized'' \citep{confirmations}.  Blockchains that rely on committee-based-consensus can achieve instant finality in the following sense.  If the system \emph{never} elects a committee with more than a $\bp$-fraction of malicious members, then as soon as a committee certifies a block, that block can be considered final, and will never be forked away.  Thus committee-based consensus protocols need to ensure that the probability a malicious committee is elected is so small, that even if the chain runs for years, there will \emph{never} be a committee with more than a $\bp$-fraction of malicious members.

Fourth, it can eliminate the need for ``slashing'' penalties.  In many traditional PoS systems (e.g. Ethereum 2.0), block producer candidates need to stake their tokens by locking them in a smart contract, and this stake is held as a bond against misbehavior.  If a block producer engages in (provable) misbehavior, their stake can be confiscated (``slashed'').  In committee-based consensus, if a small minority of the committee misbehaves, they cannot adversely affect the system, and voters (having noticed this misbehavior) will not elect them again.  For this reason, some systems (like Algorand, EOS and Tron) do not have slashing penalties.  On the other hand, Cosmos, which uses committee-based consensus \emph{does} include slashing penalties.

Finally, having a distinct separation between stakeholders and block producers allows specialization, and thus block producers in systems using committee-based consensus, may have better hardware and software infrastructure which would lead to lower latency and faster block times.

Of course, these advantages hinge on the system's ability to consistently elect an honest majority of committee members. This then raises the need to dive into the committee election mechanism.

\subsubsection*{Approval Voting}

Committee-based consensus protocols can vary on several dimensions, but we focus on \emph{how} the committee is selected.  The selection process is independent of many other features of the blockchain, e.g. the actual consensus protocol employed by the elected committee, or how data is stored and processed on the blockchain.
In this work, we focus on \emph{approval voting}, which is the selection mechanism employed by EOS and Telos, and is a strict generalization of the other common voting systems.  Approval voting is also used in other blockchain systems (outside of committee-selection), for example MakerDAO uses approval voting in its governance module\footnote{\url{https://docs.makerdao.com/smart-contract-modules/governance-module}}.

In approval voting, voters ``approve'' of a collection of candidates, and the candidates with the most approvals are elected to the committee \citep{BF07}. This is fundamentally different from traditional voting schemes, where voting for two candidates means splitting your vote.  In approval voting, if a voter votes for two (or more) candidates, each receives the same ``approval'' as if the voter only voted for one candidate.

For example, the Cosmos blockchain uses a traditional (single-vote) mechanism to elect a committee of 150 block producers\footnote{The documentation suggests 125, \citep{cosmosvalidators}, but this seems to have been increased to 150}.  By contrast, EOS uses approval voting to elect a committee of $21$ block producers.  Although Cosmos and EOS vary on several other dimensions \citep{cosmosvseos}, the committee selection mechanism is essentially independent of all these other variables.  Since Cosmos could be  modified to use approval voting, and EOS could be  modified to use a single-vote mechanism, designing the most efficient committee-based consensus protocols requires analyzing the characteristics of these mechanisms in the blockchain setting.

\setlength{\parindent}{0pt}

\section{Proofs}

\subsection{Posterior probabilities}
\label{app:posterior}

\begin{lemma}
	\label{lem:posterior}
	Let $c$ be a producer with \textit{a priori} probability to be honest $p$, suppose a voter receives a signal
	$$
	s^* = \left\{ \begin{array}{l} p_h+\eps \mbox{ if Producer $j$ is honest} \\ p_m+\eps \mbox{ if Producer $j$ is malicious } \end{array} \right.
	$$
	with $\eps \sim \mathcal{N} (0,\sigma^2)$, then
    \begin{enumerate}[label=(\roman*)]
      \item $\displaystyle{\Pr \inbrak{ c = H \suchthat s^* } = \frac{ 1 }{1 + \frac{1-p}{p}e^{\frac{(s^*-p_h)^2 - (s^*-p_m)^2}{2 \sigma^2} } }}$
      \item $\displaystyle{f_{s|H}(x) = \frac{\sigma}{\sqrt{2\pi}x(1-x)(p_h-p_m)} \cdot e^{-\inparen{\frac{ (p_h-p_m)^2 + 2 \sigma^2 \log \inparen{ \frac{p(1-x)}{(1-p)x} } }{2\sqrt{2} \sigma (p_h-p_m)} }^2 }}$
    \end{enumerate}
\end{lemma}
One useful implication of Lemma \ref{lem:posterior} is that we
can interchangeably talk about the voters considering the probabilities of producers to be honest conditioned on their signals instead of the original signals. Meaning, the model facilitates comparisons with Bayesian posteriors. In particular conditioned on $s^*_{ij}$, Lemma~\ref{lem:posterior} shows that the probability that producer $c_j$ is honest is 
\begin{equation}
s_{ij} \defined \Pr[c_j=H | s^*_{ij}]=\frac{ 1 }{1 + \frac{1-p_j}{p_j}e^{\frac{(s^*_{ij}-p_h)^2 - (s^*_{ij}-p_m)^2}{2 {\sigma_{ij}}^2} } }.
\end{equation}

\begin{proof}[Proof of Lemma~\ref{lem:posterior}]
Part (i):
	Let $f_{\eps}(x)$ denote the PDF of $\eps$.
	Bayes' Theorem says
	\begin{align}
		\Pr \inbrak{ c = H \suchthat s^* } 	&= \frac{\Pr \inbrak{c=H} \Pr \inbrak{ s^* \suchthat c = H } }{\Pr \inbrak{s^*} }  \\
											&= \frac{ \Pr \inbrak{c=H} \Pr \inbrak{ s^* \suchthat c = H } }{ \inparen{ \Pr \inbrak{c=H} \Pr \inbrak{s^* \suchthat c=H} + \Pr \inbrak{c=M} \Pr \inbrak{s^* \suchthat c=M} } } \\
											&=\frac{1}{1+\frac{\Pr \inbrak{c=M}}{\Pr \inbrak{c=H}}\frac{\Pr \inbrak{s^* \suchthat c=M}}{\Pr \inbrak{s^* \suchthat c=H}}}\\
											&= \frac{ 1 }{ 1 + \frac{1-p}{p} \frac{f_\eps(s^*-p_m)}{f_\eps(s^*-p_h)} } \\
	\end{align}

	Now,
	\begin{equation}
		\label{eqn:normal}
		f_{\eps}(x) = \frac{1}{\sigma \sqrt{2 \pi}} e^{-\frac{x^2}{2\sigma^2}}.
	\end{equation}

	Thus
	\begin{align}
		\Pr \inbrak{ c = H \suchthat s^* } = \frac{ 1 }{1 + \frac{1-p}{p} e^{\frac{(s^*-p_h)^2 - (s^*-p_m)^2}{2 \sigma^2} } }
	\end{align}
Part (ii):
Let
	\begin{equation}
		h\inparen{s^*} \defined \frac{ 1 }{1 + \frac{1-p}{p}e^{\frac{(s^*-p_h)^2 - (s^*-p_m)^2}{2 \sigma^2} } }
	\end{equation}
	then $s = h(s^*)$, by Lemma~\ref{lem:posterior}.

	Since $h(\cdot)$ is strictly increasing, the cumulative 
	density function satisfies

	\begin{equation}
		F_{s}(x) = F_{s^*}\inparen{ h^{-1}(x) }
	\end{equation}
	and the conditional cumulative distribution function satisfies
	\begin{align}
		F_{s|H}(x) &= F_{s^*|H}\inparen{h^{-1}(x)}
	\end{align}
	
	Thus it suffices to calculate $h^{-1}(\cdot)$.	
	\begin{align*}
		q &= \frac{ 1 }{1 + \frac{1-p}{p}e^{\frac{(s^*-p_h)^2 - (s^*-p_m)^2}{2 \sigma^2} } } \\
		&\Updownarrow \nonumber \\
		\frac{1}{q} &= 1 + \frac{1-p}{p}e^{\frac{(s^*-p_h)^2 - (s^*-p_m)^2}{2 \sigma^2} } \\
		&\Updownarrow \nonumber \\ 
		\frac{p}{1-p} \inparen{\frac{1}{q} - 1} &= e^{\frac{(s^*-p_h)^2 - (s^*-p_m)^2}{2 \sigma^2} } \\
		&\Updownarrow \nonumber \\ 
		\frac{p(1-q)}{(1-p)q} &= e^{\frac{(s^*-p_h)^2 - (s^*-p_m)^2}{2 \sigma^2} } \\
		&\Updownarrow \nonumber \\ 
		\log \inparen{ \frac{p(1-q)}{(1-p)q} } &= \frac{(s^*-p_h)^2 - (s^*-p_m)^2}{2 \sigma^2} \\
		&\Updownarrow \nonumber \\ 
		2 \sigma^2 \log \inparen{ \frac{p(1-q)}{(1-p)q} } &= (s^*-p_h)^2 - (s^*-p_m)^2\\
		&\Updownarrow \nonumber \\ 
		2 \sigma^2 \log \inparen{ \frac{p(1-q)}{(1-p)q} } &= 2(p_m - p_h)s^* + p_h^2 - p_m^2 \\
		&\Updownarrow \nonumber \\ 
		s^* &= \frac{ p_h^2 - p_m^2 - 2 \sigma^2 \log \inparen{ \frac{p(1-q)}{(1-p)q} } }{2(p_h-p_m)}
	\end{align*}

	Thus
	\begin{equation}
	    \label{eqn:hinv}
		h^{-1}(q) = \frac{ p_h^2 - p_m^2 - 2 \sigma^2 \log \inparen{ \frac{p(1-q)}{(1-p)q} } }{2(p_h-p_m)}
	\end{equation}

	Since $F_{s^*|H}(x) = F_{\eps}(s^* - p_h)$, and
	\begin{equation}
		f_{\eps}(x) \defined \frac{1}{\sigma \sqrt{2 \pi}} e^{-\frac{x^2}{2\sigma^2}},
	\end{equation}

	We have
	\begin{align}
		f_{s|H}(x) &= \frac{df}{dx} F_{s^*|H}(h^{-1}(x)-p_h) \nonumber \\
					&= f_{s^*|H}(h^{-1}(x)-p_h) \cdot \frac{df}{dx} h^{-1}(x) \label{eqn:chainrule}
	\end{align}
	Now

	\begin{align}
		f_{s^*|H}(h^{-1}(x)) 
					&= f_{\eps}\inparen{ h^{-1}(x) - p_h} \nonumber\\
					&= \frac{1}{\sigma \sqrt{2\pi}} e^{-\frac{ \inparen{ h^{-1}(x) - p_h }^2 }{2 \sigma^2} } \nonumber\\
					&= \frac{1}{\sigma \sqrt{2\pi}} e^{-\frac{ \inparen{\frac{ -(p_h-p_m)^2 - 2 \sigma^2 \log \inparen{ \frac{p(1-x)}{(1-p)x} } }{2(p_h-p_m)} }^2 }{2 \sigma^2} } \nonumber \\
					&= \frac{1}{\sigma \sqrt{2\pi}} e^{-\inparen{\frac{ (p_h-p_m)^2 + 2 \sigma^2 \log \inparen{ \frac{p(1-x)}{(1-p)x} } }{2\sqrt{2} \sigma (p_h-p_m)} }^2 } \label{eqn:fs}
	\end{align}
	and
	\begin{align}
		\frac{d}{dx} h^{-1}(x) &= \frac{d}{dx} \frac{ p_h^2 - p_m^2 - 2 \sigma^2 \log \inparen{ \frac{p(1-x)}{(1-p)x} } }{2(p_h-p_m)} \nonumber\\
								&= -\frac{\sigma^2}{p_h-p_m} \frac{d}{dx} \log \inparen{ \frac{p(1-x)}{(1-p)x} } \nonumber \\
								&= \frac{\sigma^2}{p_h-p_m} \frac{1}{x(1-x)} \label{eqn:dh}
	\end{align}

	Thus by Equations~\ref{eqn:chainrule}, \ref{eqn:fs} and \ref{eqn:dh}
	\begin{align}
		f_{s|H}(x) &= \frac{1}{\sigma \sqrt{2\pi}} e^{-\inparen{\frac{ (p_h-p_m)^2 + 2 \sigma^2 \log \inparen{ \frac{p(1-x)}{(1-p)x} } }{2\sqrt{2} \sigma (p_h-p_m)} }^2 } \cdot \frac{-\sigma^2}{x(x-1)(p_h-p_m)} \nonumber\\
					&= \frac{\sigma}{\sqrt{2\pi}x(1-x)(p_h-p_m)} \cdot e^{-\inparen{\frac{ (p_h-p_m)^2 + 2 \sigma^2 \log \inparen{ \frac{p(1-x)}{(1-p)x} } }{2\sqrt{2} \sigma (p_h-p_m)} }^2 } \label{eqn:fsh}
	\end{align}
	Similarly
	\begin{align}
		f_{s|M}(x) &= \frac{\sigma}{\sqrt{2\pi}x(1-x)(p_h-p_m)} \cdot e^{-\inparen{\frac{ (p_h-p_m)^2 - 2 \sigma^2 \log \inparen{ \frac{p(1-x)}{(1-p)x} } }{2\sqrt{2} \sigma (p_h-p_m)} }^2 } \label{eqn:fsm}
	\end{align}
\end{proof}

\begin{figure}[H]
	\begin{center}
		\includegraphics[width=.5\textwidth]{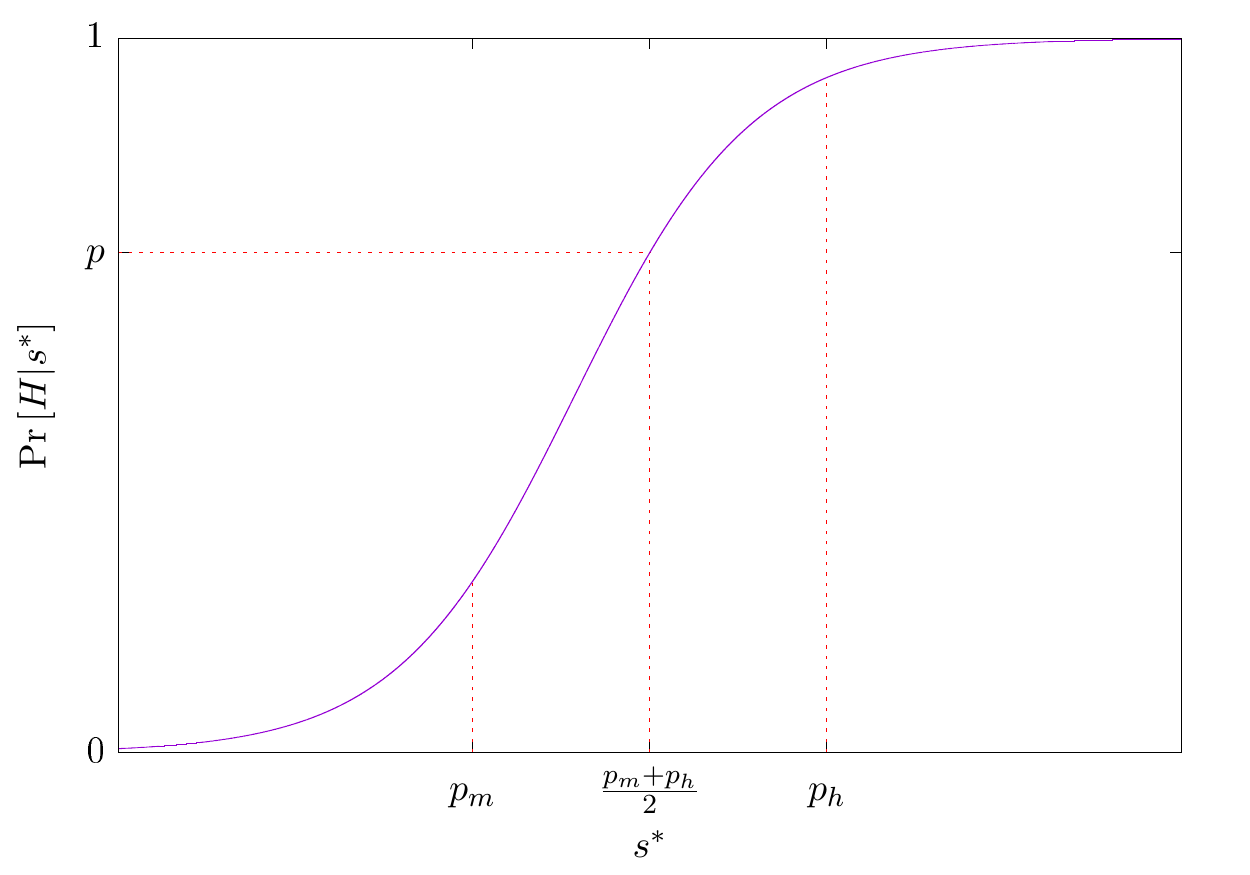}
		\caption{The probability a producer is honest, conditioned on a single voter's received signal, $s^*$}
	\end{center}
\end{figure}

\begin{figure}[H]
	\begin{center}
		\includegraphics[width=.5\textwidth]{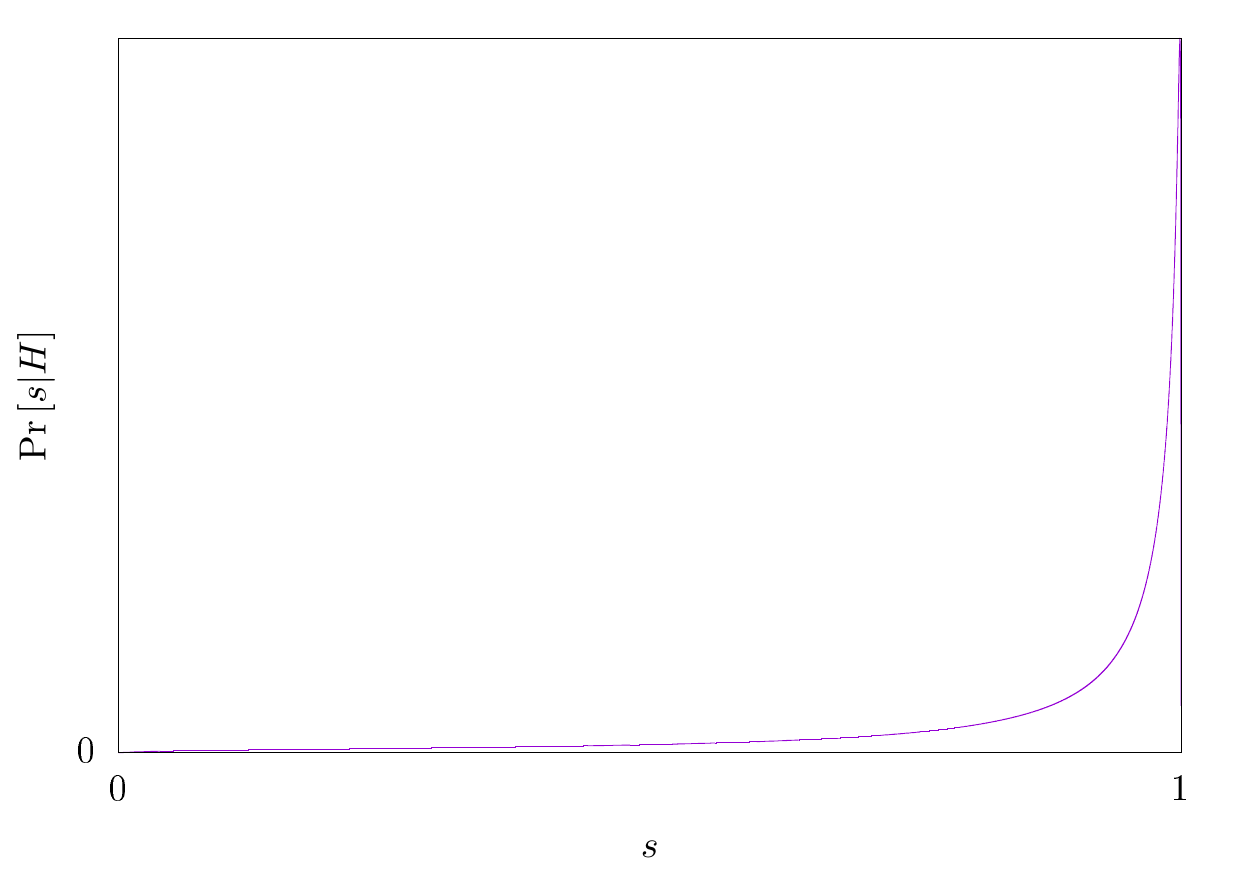}
		\caption{The distribution of the posterior probability, $s$, conditioned for an honest producer.}
	\end{center}
\end{figure}

\begin{lemma}
	\label{lem:posterior_multiple}
	Let $c_j$ be a producer with \textit{a priori} probability to be honest $p_j$, suppose voter $v_i$ receives a signal $(s^*_{ij})$ about producers $c_j$ honesty as in equation~(\ref{eqn:signal}).
	$$
	s^*_{ij} = \left\{ \begin{array}{l} p_h+\eps_{ij} \mbox{ if Producer $j$ is honest} \\ p_m+\eps_{ij} \mbox{ if Producer $j$ is malicious } \end{array} \right.
	$$
	with $\eps_{ij} \sim \mathcal{N} (0,\sigma_{ij}^2)$, then
    $$
    \Pr \inbrak{ c_j = H \suchthat s^*_{1j},\ldots,s^*_{nj} } = \frac{ 1 }{1 + \frac{1-p_j}{p_j}e^{\frac{\sum_{i=1}^n {\left[(s^*_{ij}-p_h)^2 - (s^*_{ij}-p_m)^2\right]} }{2 \sigma_{ij}^2} } }.
    $$
\end{lemma}

\begin{proof}[Proof of Lemma~\ref{lem:posterior_multiple}]
	\begin{align}
		\Pr \inbrak{ c_j = H \suchthat s^*_{1j},\ldots,s^*_{nj} } 	&= \frac{\Pr \inbrak{c_j=H} \Pr \inbrak{ s^*_{1j},\ldots,s^*_{nj} \suchthat c_j = H } }{\Pr \inbrak{s^*_{1j},\ldots,s^*_{nj}} }  \\
											&= \frac{ \Pr \inbrak{c_j=H} \Pr \inbrak{ s^*_{1j},\ldots,s^*_{nj} \suchthat c_j = H } }{ \inparen{ \Pr \inbrak{c_j=H} \Pr \inbrak{s^*_{1j},\ldots,s^*_{nj} \suchthat c_j=H} + \Pr \inbrak{c_j=M} \Pr \inbrak{s^*_{1j},\ldots,s^*_{nj} \suchthat c_j=M} } } \\
											&=\frac{1}{1+\frac{\Pr \inbrak{c_j=M}}{\Pr \inbrak{c_j=H}}\frac{\Pr \inbrak{\{s^*_{1j},\ldots,s^*_{nj} \suchthat c_j=M}}{\Pr \inbrak{\{s^*_{1j},\ldots,s^*_{nj} \suchthat c_j=H}}}\\
											&=\frac{1}{1+\frac{\Pr \inbrak{c_j=M}}{\Pr \inbrak{c_j=H}}\frac{\prod_{i=1}^n{\Pr \inbrak{s^*_{ij} \suchthat c_j=M}}}{\prod_{i=1}^n{\Pr \inbrak{s^*_{ij} \suchthat c_j=H}}}}\\
											&= \frac{ 1 }{ 1 + \frac{1-p_j}{p_j} \frac{\prod_{i=1}^n{f_{\eps_{ij}}(s^*_{ij}-p_m)}}{\prod_{i=1}^n{f_{\eps_{ij}}(s^*_{ij}-p_h)}} } \\
	\end{align}
Now,
	\begin{equation}
		f_{\eps_{ij}}(x) = \frac{1}{\sigma_{ij} \sqrt{2 \pi}} e^{-\frac{x^2}{2\sigma_{ij}^2}}.
	\end{equation}

	Thus
	\begin{align}
		\Pr \inbrak{ c_j = H \suchthat s^*_{1j},\ldots,s^*_{nj} } = \frac{ 1 }{1 + \frac{1-p_j}{p_j} e^{\frac{\sum_{i=1}^n{(s^*_{ij}-p_m)-(s^*_{ij}-p_h)}}{2\sigma_{ij}^2} } }.
	\end{align}
	
\end{proof}

\subsection{Proofs for Section \ref{sec:results}}

\begin{proof}[Proof of Theorem~\ref{thm:success}]
We condition on the number of honest and dishonest producers.
There are $m$ producers, and each is honest with probability $p$.
Suppose there are $a$ honest producers and $b \defined m - a$ dishonest producers.
(Note that $a \sim \Bin\inparen{ m, p }$).

The committee is \emph{honest} if at least $\inceil{\inparen{1-\bp}\cdot k}$ of the elected block producers are honest.
\begin{claim}
\label{claim:cmmittee_order_stats}
 The committee is honest if and only if the $\inceil{\inparen{1-\bp}\cdot k}$-st top-ranked honest producer has more votes than the $\inceil{\bp \cdot k}$-th ranked dishonest producer
\end{claim}
Let $X^h_1,\ldots,X^h_a$ denote the number of votes received by an honest producer $j$ and $X^h_{a,1},\ldots,X^h_{a,a}$ their order statistics. Similarly, let $X^m_1,\ldots,X^m_b$ denote the number of votes received by a malicious producer $j$ and $X^m_{b,1},\ldots,X^m_{b,b}$ their order statistics. With this notation Claim~\ref{claim:cmmittee_order_stats} becomes:
\begin{equation}
	\ord{X^h}{a-\inceil{\inparen{1-\bp}\cdot k}+1}{a} > \ord{X^m}{b-\inceil{\bp \cdot k}+1}{b}
\end{equation}

We define $X \defined \ord{X^h}{a-\inceil{\inparen{1-\bp}\cdot k}+1}{a}$ and $Y \defined \ord{X^m}{b-\inceil{\bp \cdot k}+1}{b}$. 

If $a<\inceil{\inparen{1-\bp}\cdot k}$ then the success probability is zero. Also if $a>m-\inceil{\bp \cdot k}$ then $b<\inceil{\bp \cdot k}$ so the success probability is $1$ hence:
\begin{equation}
    \label{eq:success_prob_final}
    \Pr\inbrak{\mathbb{T}=H}=\sum_{a=m-\inceil{\bp \cdot k}+1}^m{1\cdot\Pr\inbrak{a=a}}+\sum_{a=\inceil{\inparen{1-\bp}\cdot k}}^{m-\inceil{\bp \cdot k}}{\Pr\inbrak{X>Y|a=a}\Pr\inbrak{a=a}}
\end{equation}
We have:
\begin{equation}
\label{eq:binom_honest}
    \Pr\inbrak{a=a}\sim \Bin\inparen{ m, p }=\binom{m}{a}p^a(1-p)^{m-a}
\end{equation}

For any discrete, independent random variables, $X$ and $Y$ 
\begin{equation}
	\Pr [X > Y] = \sum_{x} \Pr[ X = x ] \Pr[ Y < x ]
\end{equation} so together with  Theorems~\ref{thm:order_pdf},\ref{thm:order_cdf} we have:
\begin{align}
	\Pr \inbrak{ X > Y} 	&= \sum_{x=0}^n \Pr [X = x] \Pr [Y < x] \\
						 	&= \sum_{x=0}^n f_X(x)(F_Y(x)-f_Y(x))\\				&= \sum_{x=0}^n \Biggl[ \left( \sum_{j=0}^{\inceil{\inparen{1-\bp}\cdot k}-1} \binom{a}{j} \left( {\left( 1 - F^h(x) \right)}^j \inparen{F^h(x)}^{a-j} - \inparen{ 1 - F^h(x) + f^h(x) }^j \inparen{ F^h(x) - f^h(x) }^{a-j} \right) \right) \Biggr. \\ 
							&\Biggl. \left( \sum_{j=0}^{\inceil{\bp \cdot k}-1} \binom{b}{j} \inparen{ 1- F^m(x) +f^m(x) }^j \inparen{F^M(x)-f^m(x)}^{b-j} \right) \Biggr] 
\end{align}

Plugging the last equation and Equation~\ref{eq:binom_honest} to Equation~\ref{eq:success_prob_final} gives the result.
\end{proof}

\begin{proof}[Proof of Theorem~\ref{thm:voting_distribution}]
   If the probability that voter $i$ votes for candidate $j$ when candidate $j$ is honest (resp. dishonest) is $p_i^h$ (resp. $p_i^m$), then the total number of votes for candidate $j$ is distributed as a Poisson Binomial with parameters $p_1^h,\ldots,p_n^h$ (resp. $p_1^m,\ldots,p_n^m$).
  
  Then applying Equation~\ref{eqn:PBDcount} in Definition~\ref{def:PBD} gives the result.
\end{proof}

\begin{proof}[Proof of Proposition~\ref{prop:threshold}]
The number of votes received by an honest producer is distributed according to a Poisson random 
	variable with parameters $p_1^h,\ldots,p_n^h$, where 
	\begin{align}
		p_i^h   &= 1 - F_{s|H}(z_i)\\
		        &= 1 - \Phi\inparen{ \frac{h^{-1}(z_i) - p_h }{\sigma_i} }
	\end{align}	
	where $\Phi$ is the CDF of the standard normal distribution and $h^{-1}(\cdot)$ is defined in Equation~\ref{eqn:hinv}.
	Similarly, the number of of votes received by a dishonest producer is distributed according to a Poisson random 
	variable with $p_1^m,\ldots,p_n^m$, where
	\begin{align}
		p_i^m   &= 1 - F_{s|M}(z_i)\\
		        &= 1 - \Phi\inparen{ \frac{h^{-1}(z_i) - p_m }{\sigma_i} }
	\end{align}
\end{proof}

\begin{proof}[Proof of Proposition~\ref{prop:cardinal}]
	When voters follow the cardinal strategy, for voter $i$ to vote for candidate $j$, its signal needs to be among the top $z_i$ signals, meaning:
	\begin{equation}
	    p_i=\Pr \inbrak{s_{ij}\text{ is in the top }z_i\text{ signals} }
	\end{equation}
	Let $S^i_{m,1},\ldots,S^i_{m,m}$ denote that order statistics of $s_{i1},\ldots,s_{im}$. Then for candidate $j$ to be chosen by voter $i$ it needs to hold that $s_{ij}\geq S^i_{m,z_i}$ so:
	\begin{equation}
	\label{eq:cardinal_prob}
	    p_i=\Pr \inbrak{s_{ij}\geq S^i_{m,z_i}}=\Pr \inbrak{s_{i}\geq S^i_{m,z_i}}
	\end{equation}
	with the last equality since we assumed $\sigma_{ij}=\sigma_i$. Let $X_i=s_i$ and $Y_i=S^i_{m,z_i}$ then equation~\ref{eq:cardinal_prob} becomes:
	\begin{equation}
	\label{eq:cardinal_int}
	    p_i=\Pr \inbrak{X_i\geq Y_i}=\int_{-\infty}^{\infty}{\Pr\inbrak{Y_i\leq x}\Pr\inbrak{X_i=x}\,dx}=\int_{-\infty}^{\infty}{F_{Y_i}(x)f_{X_i}(x)\,dx}
	\end{equation}
	As in the Proof of Theorem~\ref{thm:success}, we condition on the number of honest producers: $a$. $a\sim \Bin\inparen{ m, p }$ where $p$ is the probability of a producer to be honest. The real valued random variables $(s_{ij})_{1\leq j\leq m}$ are drawn from two populations. Lemma $2$ provides the PDF and CDF of honest producers $f_{s_i|H},F_{s_i|H}$ and malicious producers $f_{s_i|M},F_{s_i|M}$. W.L.O.G, suppose  that $F_{s_{ij}}=F_{s_i|H}$ for all $1\leq j \leq a$ and $F_{s_{ij}}=F_{s_i|M}$ for all $a+1\leq j \leq m$. By Theorem~\ref{thm:two_populations} we have that
	\begin{equation}
	\label{eq:cardinal_order_stats_cdf}
	\begin{aligned}
	    F_{Y_i}(x)&=\sum_{l=z_i}^{m}{\frac{(a!(m-a)!)^2}{z_i!(m-z_i)!}}\\
	    &\sum_{\substack{0\leq \lambda_1\leq z_i\\0\leq \lambda_2\leq m-z_i\\ \lambda_1+\lambda_2=a}}{\left[\binom{z_i}{\lambda_1}\inparen{F_{s_i|H}}^{\lambda_1}\inparen{F_{s_i|M}}^{z_i-\lambda_1}\binom{m-z_i}{\lambda_2}\inparen{1-F_{s_i|H}}^{\lambda_2}\inparen{1-F_{s_i|M}}^{m-z_i-\lambda_2}\right]}
	\end{aligned}
	\end{equation}
	So finally Equation~\ref{eq:cardinal_int} becomes:
	\begin{equation}
	\label{eq:cardinal_final}
	    p_i=\sum_{a=0}^m{p^a(1-p)^{m-a}}  \int_{-\infty}^{\infty}{F_{Y_i(x)}f_{s_i}(x)\,dx},
	\end{equation}
	where $F_{Y_i}$ is given by Equation~\ref{eq:cardinal_order_stats_cdf}. If the candidate is honest then $p_i^h$ is given by Equation~\ref{eq:cardinal_final} with $f_{s_i}=f_{s_i|H}$ derived in Lemma~\ref{lem:posterior}. Similarly, if the candidate is malicious then $p_i^m$ is given by Equation~\ref{eq:cardinal_final} with $f_{s_i}=f_{s_i|M}$.
\end{proof}

\begin{proof}[Proof of Lemma~\ref{lem:poisson_maximality}]
Follows immediately from Lemma \ref{lem:symmetry},\ref{lem:monotonicity} and stochastic order definition~\ref{def:ordering}.
\end{proof}

\begin{proof}[Proof of Proposition~\ref{prop:single}]
Suppose that voter $v_1$ receives conditioned signals $s_{j}$ on candidate $c_j$ and they are sorted so that $s_{1}\geq \cdots \geq s_{m}$. Let $\cset_{v_1}=\{c_{j_1},\ldots,c_{j_z}\}$ be voter $v_1$ strategy then  the candidates on the chosen committee are $\mathbb{T}=\{c_{j_1},\ldots,c_{j_z},c_{j_{z+1}},\ldots,c_{j_k}\}$ where $c_{j_{z+1}},\ldots,c_{j_k}$ are filled adversarially. Let $X$ be the random variable that is the number of honest producers on the chosen committee. Then $X$ is a Poisson Binomial with probabilities $(s_{j_1},\ldots,s_{j_z},s_{j_{z+1}},\ldots,s_{j_k})$ and so $\Pr[\mathbb{T}=H]=\Pr[X>(1-\bp)\cdot k]$. Define $X_P$ with $P=\{s_{j_1},\ldots,s_{j_l}\} \subseteq \{s_1,\ldots,s_m\}$ to be the Poisson Binomial with parameters  $(s_{j_1},\ldots,s_{j_l})$ then by Lemma \ref{lem:poisson_maximality} we have (with $x=(1-\bp)\cdot k$) that 
\begin{equation}
    \label{eq:pois_max_1}
    \underset{P\subseteq \{s_1,\ldots,s_m\}}{\operatorname{argmax}} {\Pr [X_P> (1-\bp)\cdot k]}=\{s_1,\ldots,s_k\}
\end{equation}

So by definition the optimal strategy for $v_1$ is achieved by setting $z=k$ and $\cset_{v_1}=\{c_{1},\ldots,c_{k}\}$. That is, choosing the top $z=k$ candidates which is the Cardinal strategy with $z=k$. 
\end{proof}

\begin{proof}[Proof of Proposition~\ref{prop:threshold_suboptimal}]
	It suffices to consider the single voter case. Let $z\in (0,1)$	and suppose voter $v_1$ receives ordered, conditioned signals $s_j$, follows the threshold strategy and votes for all candidates $s_j>z$.
	Let $m_z \in [0,\ldots,m]$ denote the number of candidates that receive a vote.
	
	If $m_z\neq k$ then this strategy is not optimal by equation~\ref{eq:pois_max_1}. 
	To show that the Threshold strategy is not optimal, it remains to show that $\Pr[m_z\neq k]>0$. 

	Since $m_z$ is distributed as a Poisson Binomial with parameters $p_1,\ldots,p_m$ with $p_j \defined \Pr \inbrak{ s_j > z }$ and $p_j>0,\forall j\in[1,\ldots,m]$, 	
	we have that $\Pr \inbrak{ m_z = i } > 0$ for all $i \in [0,\ldots,m]$, which means that $\Pr \inbrak{ m_z = k } < 1$.
\end{proof}

\begin{proof}[Proof of Proposition~\ref{prop:multiple}]
Suppose that the resulting posterior probabilities of honesty for the candidates using Lemma~\ref{lem:posterior_multiple} are $s_j$ and they are sorted so that $s_{1}\geq \cdots \geq s_{m}$. Let the elected candidates on the committee be $\mathbb{T}=\{c_{j_1},\ldots,,c_{j_k}\}$. Let $X$ be the random variable that is the number of honest producers on the chosen committee. Then $X$ is a Poisson Binomial with probabilities $(s_{j_1},\ldots,s_{j_k})$ and so $\Pr[\mathbb{T}=H | \mathbf{s}_i]=\Pr[X>(1-\bp)\cdot k]$. Define $X_P$ with $P=\{s_{j_1},\ldots,s_{j_l}\} \subseteq \{s_1,\ldots,s_m\}$ to be the Poisson Binomial with parameters  $(s_{j_1},\ldots,s_{j_l})$ then by Lemma \ref{lem:poisson_maximality} we have (with $x=(1-\bp)\cdot k$) that 
\begin{equation}
    \label{eq:pois_max_2}
    \underset{P\subseteq \{s_1,\ldots,s_m\}}{\operatorname{argmax}} {\Pr [X_P>(1-\bp)\cdot k]}=\{s_1,\ldots,s_k\}
\end{equation}

If we set $z=k$ and $\cset_{v_i}=\{c_{1},\ldots,c_{k}\}$. That is, choosing the top $z=k$ candidates based on the posteriors $s_j$ we achieve the maximum on the RHS of equation~(\ref{eq:pois_max_2}).
\end{proof}

\begin{proof}[Proof of Proposition~\ref{prop:cardinal_suboptimal}]
Proposition \ref{prop:single} shows that for $z\neq k$ the Cardinal strategy may be suboptimal. 

Consider a setting with two voters, where $p_m = p_h$.
When $p_m = p_h$, all the signals are uninformative, i.e., $s_{ij} = p$ for all $i \in [n]$, $j \in [m]$.

Now, suppose both voters follow the cardinal strategy with thresholds $z_0,z_1$.
In this case, voter $i$ will vote (randomly) for $z_i$ candidates.

Since there are only two voters, every candidate receives $0,1$ or $2$ votes.
Let $X_0,X_1,X_2$ denote the subsets producers that receive $0,1$ and $2$ votes respectively.
Let $\cH$ and $\cM$ denote the set of honest and dishonest producers.

Thus $X_0 \du X_1 \du X_2 = [m] = \cH \du \cM$, where $\du$ denotes the disjoint union of sets.

First, note that since there are two voters, 
\begin{equation}
	\left| X_1 \right| + 2\left| X_2 \right| = z_0 + z_1
\end{equation}

Since $\left| X_2 \right| \le \min(z_0,z_1)$, we have $\left| X_1 \cup X_2 \right| \ge \max(z_0,z_1)$, 
which means that when $\max(z_0,z_1) \ge k$ (i.e., either voter votes for $k$ candidates) no candidates from $X_0$ will ever make it to the committee.

To show that the threshold $z_i = k$ strategy is suboptimal, it suffices to consider strategies with $\max(z_0,z_1) \ge k$.

Since we assume that ties are broken in a worst-case fashion,
the committee will be dishonest if and only if there are $x$ dishonest candidates in $X_2$ and there are at least $t+1-x$ dishonest candidates in $X_1$ and $|X_2| \le k-(t+1-x)$.
because in this case, the adversary can choose $k-(t+1-x)$ dishonest candidates from $X_1$ to fill the committee.

In other words, the committee will be dishonest with probability

\begin{align}
	\label{eqn:dishonest_com}
	\Pr \inbrak{ \mathbb{T} \ne H } &= 
		\sum_{a = 0}^{\min(z_0,z_1)} \Pr [ |X_2| = a ] \sum_{x=0}^a \Pr \inbrak{ |X_2 \cap \cM| = x \suchthat |X_2| = a } \Pr \inbrak{ |X_1 \cap \cM| \ge t+1-x }
\end{align}

Note that if we did \emph{not} assume $\min(z_0,z_1) \ge k$, some candidates from $X_0$ could make it onto the committee, and Equation~\ref{eqn:dishonest_com} would have extra terms.

Now $|X_2|$ is distributed as a hypergeometric random variable with parameters $\inparen{z_0,z_1,m}$ (i.e., $z_1$ draws from a population of size $m$ with $z_0$ ``distinguished'' items), 
and $|X_1|$ is distributed as a $z_0 + z_1 - 2|X_2|$.

Thus we have
\begin{equation}
	\Pr \inbrak{ |X_2| = a } = \frac{ \binom{z_0}{a} \binom{m-z_0}{z_1-a} }{\binom{m}{z_1}}
\end{equation}

Since each producer is dishonest independently with probability $p$, the number of dishonest producers in $X_2$ is binomial random variable with parameters $a,p$.
\begin{align}
	\Pr \inbrak{ |X_2 \cap \cM| = x \suchthat |X_2| = a } 	&= \Pr \inbrak{ \Bin(a,p) = x } \\
\end{align}
and
\begin{align}
	\Pr \inbrak{ |X_1 \cap \cM| = x \suchthat |X_2| = a } 	&= \Pr \inbrak{ |X_1 \cap \cM| = x \suchthat |X_1| = z_0+z_1 -2a  }\\
															&= \Pr \inbrak{ \Bin(z_0+z_1-2a,p) = x } \\
\end{align}

Equation~\ref{eqn:dishonest_com} becomes

\begin{align}
	\label{eqn:dishonest_com2}
	\Pr \inbrak{ \mathbb{T} \ne H } &= \sum_{a = 0}^{\min(z_0,z_1)} \Pr \inbrak{ |X_2| = a } \cdot \sum_{x=0}^a \Pr \inbrak{ \Bin(a,p) = x } \cdot \Pr \inbrak{ \Bin(z_0+z_1-2a, p) \ge t+1-x } \\
									&= \sum_{a = 0}^{\min(z_0,z_1)} \Pr \inbrak{ |X_2| = a } \cdot \Pr \inbrak{ \Bin(z_0+z_1-a, p) \ge t+1 } \nonumber \\
									&\ge \sum_{a = 0}^{\min(z_0,z_1)} \Pr \inbrak{ |X_2| = a } \cdot \Pr \inbrak{ \Bin(\max(z_0,z_1), p) \ge t+1 } \nonumber \\
									&= \Pr \inbrak{ \Bin(\max(z_0,z_1), p) \ge t+1 } \nonumber \\
\end{align}

When $z_0 = z_1 = k$ (both voters vote for $k$ producers) Equation~\ref{eqn:dishonest_com2} becomes
\begin{align}
	\label{eqn:dishonest_comk}
	&\sum_{a = 0}^{k} \Pr \inbrak{ |X_2| = a } \cdot \sum_{x=0}^a \Pr \inbrak{ \Bin(a,p) = x } \cdot \Pr \inbrak{ \Bin(2(k-a), p) \ge t+1-x } \\
	&=\sum_{a = 0}^{k} \Pr \inbrak{ |X_2| = a } \cdot \Pr \inbrak{ \Bin(2k-a, p) \ge t+1 } \\
	&>\sum_{a = 0}^{k} \Pr \inbrak{ |X_2| = a } \cdot \Pr \inbrak{ \Bin(k, p) \ge t+1 } \\
	&= \Pr \inbrak{ \Bin(k,p) \ge t + 1 }
\end{align}

When $z_0 = k$, and $z_1 = 0$ (voter 1 abstains) Equation~\ref{eqn:dishonest_com2} becomes
\begin{equation}
	\label{eqn:dishonest_com0}
	\Pr \inbrak{ \Bin(k,p) \ge t+1} = \sum_{b=t+1}^{k} \binom{k}{b} (1-p)^b p^{k-b} 
\end{equation}

Here, we see that Equation~\ref{eqn:dishonest_comk} is strictly greater than Equation~\ref{eqn:dishonest_com0}, so voter $1$ is strictly better off setting $z_1 = 0$ (voting for no producers) 
than voting for $k$ producers.
\end{proof}

\begin{proof}[Proof of Theorem~\ref{thm:approximate_optimality}]
	First, note that since $|\cH| \ge (1-\bp)\cdot k$, then if \emph{all} members of $\cH$ receive
	more votes than all dishonest producers, then the committee will be honest.
	Thus it suffices to show that all members of $\cH$ will receive more votes than all members of $\cM$ 
	with high probability.

	Let $X_j$ denote the number of votes received by producer $j$.
	Then $X_j$ is a Poisson Binomial with parameters $p_{1j},\ldots,p_{nj}$.

	Fix $j_h \in \cH$, and $j_m \in \cM$.
	Then, by assumption
	\begin{equation}
		\sum_{i=1}^n p_{ij_h} \ge n \delta + \sum_{i=1}^n p_{i j_m}
	\end{equation}

	The Chernoff bound for Poisson Binomials (Theorem~\ref{thm:CH}) shows that for all $t>0$
	\begin{equation}
		\Pr \inbrak{ X_{j_h} < \sum_{i=1}^n p_{ij_h} - t} \le e^{-\frac{2t^2}{n}}
	\end{equation}
	and similarly
	\begin{equation}
		\Pr \inbrak{ X_{j_m} > \sum_{i=1}^n p_{ij_m} + t} \le e^{-\frac{2t^2}{n}}
	\end{equation}

	If $X_{j_h} \le X_{j_m}$, then either 
	\begin{equation}
		X_{j_h} < \sum_{i=1}^n p_{ij_h} - \frac{n \delta}{2}
	\end{equation}
	or
	\begin{equation}
		X_{j_m} > \sum_{i=1}^n p_{ij_m} + \frac{n \delta}{2}.
	\end{equation}

	By a union bound, the probability that either of these events happens is bounded by 
	\begin{equation}
		 2e^{- \delta^2 n/2  }.
	\end{equation}

	Taking a union bound over all pairs $j_h \in \cH$, and $j_m \in \cM$, we have 
	the probability that \emph{all} $j_h \in \cH$ receive more votes than \emph{all} $j_m \in \cM$, 
	is at least

	\begin{equation}
		1 - 2m^2 e^{-\delta^2 n/2}.
	\end{equation}
\end{proof}

\section{Poisson Binomial Distributions}
\label{app:PBD}

\subsection{Definitions}

\begin{definition}[Poisson Binomial Distribution]
	\label{def:PBD}
	If $\inset{X_i}$ are independent Bernoulli random variables, 
	and $\Pr [ X_i = 1 ] = p_i$, then the distribution of $X \defined \sum_i X_i$ 
	is called the \emph{Poisson Binomial Distribution} with parameters $(p_1,\ldots,p_n)$.
\end{definition}

A simple counting argument shows that if $X$ has a Poisson Binomial distribution with parameters $p_1,\ldots,p_n$, 
then for any $\ell \in 0,\ldots,n$

\begin{equation}
	\label{eqn:PBDcount}
	\Pr[X=\ell]=\sum_{A\in F_{\ell}}{\prod_{i\in A}{p_i}\prod_{j\in A^c}{1-p_j}}
\end{equation} 
where $F_{\ell}$ is the set of all subsets of $\ell$ integers that can be selected from $\inset{1,\ldots,n}$.

\begin{lemma}[Symmetry of Poisson Binomial Distributions]
    \label{lem:symmetry}
	Suppose $\sigma: [n] \rightarrow [n]$ is a permutation, 
	and let $X$ be a Poisson Binomial random variable with parameters $p_1,\ldots,p_n$, and 
	$X'$ be a Poisson Binomial random variable with parameters $\sigma(p_1),\ldots,\sigma(p_n)$, 
	then for all $\ell \in 0,\ldots,n$,
	\[
		\Pr \inbrak{ X = \ell} = \Pr[ X' = \ell ]
	\]
\end{lemma}

\begin{lemma}[Monotonicity]
	\label{lem:monotonicity}
	Suppose $X$ is a Poisson Binomial random variable with parameters $p_1,\ldots,p_n$, and 
	$X'$ is a Poisson Binomial random variable with parameters $p'_1,\ldots,p'_n$, 
	satisfying $p_i \le p'_i$ for $i = 1,\ldots,n$, 
	then 
	\[
		X \le_{\mathrm{st}} X'.
	\]
\end{lemma}

\begin{proof}
	The proof follows immediately from Lemma~\ref{lem:sum} since a Bernoulli random variable with parameter $p'_i$ stochastically 
	dominates a Bernoulli random variable with parameter $p_i \le p'_i$.
\end{proof}

\subsection{Alternative characterizations of Poisson Binomial Distribution}

\begin{theorem}[Alternative characterization of a PDF of a Poisson Binomial \cite{FW10}]
    \label{thm:altPBDPDF}
	If $X$ has a Poisson Binomial Distribution with parameters $(p_1,\ldots,p_n)$, 
	then
	\begin{equation}
		\label{eqn:PBPDF}
			\Pr \inbrak{ X = t } = \frac{1}{n+1} \sum_{j=0}^n \inparen{ e^{-2 \pi i \frac{jt}{n+1} } \prod_{k=1}^n \inparen{ p_k e^{2 \pi i \frac{j}{n+1} } + (1-p_k) } }
	\end{equation}
	We use $\PBPDF_{p_1,\ldots,p_n}(x)$ to denote the PDF of a Poisson Binomial random variable with parameters $p_1,\ldots,p_n$.
\end{theorem}

\begin{theorem}[CDF of a Poisson Binomial \cite{FW10}]
    \label{thm:altPBDCDF}
	If $X$ has a Poisson Binomial Distribution with parameters $(p_1,\ldots,p_n)$, 
	then
	\begin{equation}
		\label{eqn:PBCDF}
			\Pr \inbrak{ X \ge t } = 1 - \frac{1}{n+1}\sum_{j=0}^n \inparen{ \inparen{ \sum_{k=0}^{t-1} e^{ -2 \pi i \frac{jk}{n+1} } } \prod_{\ell=1}^n p_\ell e^{ 2 \pi i \frac{j}{N+1} } + (1-p_\ell) }
	\end{equation}
	We use $\PBCDF_{p_1,\ldots,p_n}(x)$ to denote the CDF of a Poisson Binomial random variable with parameters $p_1,\ldots,p_n$.
\end{theorem}

\subsection{Concentration bounds}

\begin{theorem}[Chernoff-Hoeffding {\cite[Theorem 1.1]{DP09}}]
	\label{thm:CH}
	Let $X = \sum_{i=1}^n X_i$ is a Poisson Binomial with parameters $\inset{p_i}$, and define $\bar{p} \defined \frac{1}{n} \sum_{i=1}^n p_i$,
	then
	\begin{align}
		\Pr \inbrak{ X > n\bar{p} + t } \le e^{-2t^2/n} \\
		\Pr \inbrak{ X < n\bar{p} - t } \le e^{-2t^2/n} 
	\end{align}
\end{theorem}

Linear combinations of order statistics of Poisson Binomial RVs also satisfy a central limit theorem \cite{PT82}.

\section{Order Statistics}

\begin{definition}[Order statistics]
	Let $X_1,\ldots,X_n$ be random variables.
	Define $\ord{X}{1}{n},\ldots\ord{X}{n}{n}$ to be the \emph{order statistics} of $X_1,\ldots,X_n$, 
	to $X_1,\ldots,X_n$ in sorted order.
\end{definition}

\begin{remark}
	The random variables $\ord{X}{1}{n},\ldots,\ord{X}{n}{n}$ are \emph{dependent} even if the underlying $\inset{X_i}$ are independent, 
	and they satisfy
	\begin{equation}
		\ord{X}{1}{n} \le \cdots \le \ord{X}{n}{n}.
	\end{equation}
\end{remark}

\begin{theorem}
	\label{thm:order_cdf_continuous}
	If $\inset{X_i}$ are \emph{continuous} IID random variables, with absolutely continuous PDF $f(x)$ and CDF, $F(x)$, then
	the CDF of the $k$th order statistic from a sample of size $n$ is 
	\begin{equation}
		\Pr \inbrak{ \ord{X}{k}{n} \le x } = F_{(k,n)}(x) = \sum_{j=k}^{n} \binom{n}{j} \inparen{ F(x) }^j \inparen{1-F(x)}^{n-j}
	\end{equation}
\end{theorem}

\begin{theorem}
	\label{thm:order_pdf}
	If $\inset{X_i}$ are \emph{discrete} IID random variables, with PDF $f(x)$ and CDF, $F(x)$, then
	the PDF of the $k$th order statistic from a sample of size $n$ is 
	\begin{equation}
		\Pr \inbrak{ \ord{X}{k}{n} = x} = f_{(k,n)}(x) = \sum_{j=0}^{n-k} \binom{n}{j} \inparen{ \inparen{ 1 - F(x) }^j \inparen{F(x)}^{n-j} - \inparen{ 1 - F(x) + f(x) }^j \inparen{ F(x) - f(x) }^{n-j} }.
	\end{equation}
\end{theorem}

\begin{theorem}
	\label{thm:order_cdf}
	If $\inset{X_i}$ are \emph{discrete} IID random variables, with PDF $f(x)$ and CDF, $F(x)$, then
	the CDF of the $k$th order statistic from a sample of size $n$ is 
	\begin{equation}
		\Pr \inbrak{ \ord{X}{k}{n} \le x } = F_{(k,n)}(x) = \sum_{j=0}^{n-k} \binom{n}{j} \inparen{ 1- F(x) }^j \inparen{F(x)}^{n-j}
	\end{equation}
\end{theorem}

\begin{theorem}[Bapat-Beg Theorem {\cite[Theorem 3.1]{GKMHM08}}]
	\label{thm:bapat_beg}
	Let $X_i$, $i=1,\ldots,m$ independent, real-valued random variables with cdf $F_i(x)$ respectively. $Y_i$ the order statistics defined by sorting the values of $X_i$. Let $n,1\leq n_1 < n_2 < \cdots <n_k \leq m$ and $y_1\leq y_2 \leq \cdots \leq y_k$ the values of the arguments of the joint cdf of $\{ Y_{n_1},Y_{n_2},\ldots,Y_{n_k}\}$. Define the index vector $\mathbf{i}=(i_0,i_1,\ldots,i_{k+1})$ and the summation index set $\mathcal{I} = \{ \mathbf{i}: 0=i_0\leq i_1\leq \cdots \leq i_k\leq i_{k+1}=m,\text{ and } i_j\geq n_j \text{ for all } 1\leq j \leq k \}$. The joint cdf of the order statistics satisfies:
	\begin{equation}
		F_{Y_{n_1},\ldots,Y_{n_k}}(y_1,\ldots,y_k) = \sum_{\mathbf{i}\in \mathcal{I}}{\frac{P_{i_1,\ldots,i_k}(y_1,\ldots,y_k)}{(i_1-i_0)!(i_2-i_1)!\cdots (i_{k+1}-i_k)!}},
	\end{equation}
where $P_{i_1,\ldots,i_k}(y_1,\ldots,y_k)$ are permanents of block matrices 
\begin{equation}
		P_{i_1,\ldots,i_k}(y_1,\ldots,y_k) = \operatorname{per}\left[ [F_i(y_j)-F_i(y_{j-1})]_{(i_j-i_{j-1})\times 1} \right]_{j=1,i=1}^{j=k,i=m},
	\end{equation}
with the subscripts indicating the dimensions of blocks created by the repetition of the term in the brackets, and $F_i(y_0)=0, F_i(y_{k+1})=1$.
\end{theorem}

\begin{theorem}[{\cite[Theorem 3.2]{GKMHM08}}]
	\label{thm:two_populations}
	With notation as in Theorem \ref{thm:bapat_beg}, suppose that $F_i(x)=F(x)$ for all $1\leq i \leq n$, and $F_i(x)=G(x)$ for all $n+1\leq i \leq m$. Then: 
	\begin{equation}
		F_{Y_{n_1},\ldots,Y_{n_k}}(y_1,\ldots,y_k) = \sum_{\mathbf{i}\in \mathcal{I}}{\sum_{\boldsymbol{\lambda}}{\prod_{j=1}^{k+1}{\frac{n!(m-n)!}{\lambda_j!(i_j-i_{j-1}-\lambda_j)!}\left[F(y_j)-F(y_{j-1}) \right]^{\lambda_j}\left[G(y_j)-G(y_{j-1}) \right]^{i_j-i_{j-1}-\lambda_j}}}},
	\end{equation}
where $\boldsymbol{\lambda}=(\lambda_1,\lambda_2,\ldots ,\lambda_{k+1})$ ranges over all integer vectors such that
\begin{equation}
		\lambda_1+\lambda_2+\cdots+\lambda_{k+1} = n, \, 0\leq \lambda_j \leq i_j-i_{j-1},
	\end{equation}
and $F(y_0)=G(y_0)=0,F(y_{k+1})=G(y_{k+1})=1$.
\end{theorem}

\begin{definition}[Orderings]
    \label{def:ordering}
	Let $X$ and $Y$ be random variables with PDFs $f_X$, $f_Y$ and CDFs $F_X$, $F_Y$.
	\begin{itemize}
		\item
			\textbf{Stochastic Order:}
				\begin{equation}
					X \le_{\mathrm{st}} Y \Leftrightarrow F_Y(x) \le F_X(x) \mbox{ for all $x$ },
				\end{equation}
		\item
			\textbf{Hazard-rate Order:}
				\begin{equation}
					X \le_{\mathrm{hr}} Y \Leftrightarrow (1-F_Y(x))/(1-F_X(x)) \mbox{ is increasing in $x$ },
				\end{equation}
		\item
			\textbf{Likelihood-ratio Order:}
				\begin{equation}
					X \le_{\mathrm{lr}} Y \Leftrightarrow f_Y(x)/f_X(x) \mbox{ is increasing in $x$ },
				\end{equation}
	\end{itemize}
\end{definition}

\begin{lemma}[Summations]
	\label{lem:sum}
	If $X_1 \ge_{\mathrm{st}} Y_1$, and $X_2 \ge_{\mathrm{st}} Y_2$, then
	\[
		X_1 + X_2 \ge_{\mathrm{st}} Y_1 + Y_2
	\]	
\end{lemma}

\begin{proof}
	\begin{align*}
		\Pr \inbrak{ X_1 + X_2 \ge t } 	&= \sum_{x_1=0}^n \sum_{x_2=t-x_1}^n \Pr \inbrak{ X_1 = x_1 } \Pr\inbrak{X_2 = x_2} \\
									 	&= \sum_{x_1=0}^n \inparen{\sum_{x_2=t-x_1}^n \Pr\inbrak{X_2 = x_2}} \Pr \inbrak{ X_1 = x_1 }  \\
									 	&= \sum_{x_1=0}^n \Pr \inbrak{ X_2 \ge t-x_1 } \Pr \inbrak{ X_1 = x_1 }  \\
									 	&\ge \sum_{x_1=0}^n \Pr \inbrak{ Y_2 \ge t-x_1 } \Pr \inbrak{ X_1 = x_1 }  \\
									 	&= \sum_{x_1=0}^n \inparen{\sum_{x_2=t-x_1}^n \Pr\inbrak{Y_2 = x_2}} \Pr \inbrak{ X_1 = x_1 }  \\
									 	&= \sum_{x_2=0}^n \inparen{\sum_{x_1=t-x_2}^n \Pr\inbrak{X_1 = x_1}} \Pr \inbrak{ Y_2 = x_2 }  \\
									 	&= \sum_{x_2=0}^n \Pr \inbrak{ X_1 \ge t - x_2 } \Pr \inbrak{ Y_2 = x_2 }  \\
									 	&\ge \sum_{x_2=0}^n \Pr \inbrak{ Y_1 \ge t - x_2 } \Pr \inbrak{ Y_2 = x_2 }  \\
									 	&= \sum_{x_2=0}^n \inparen{\sum_{x_1=t-x_2}^n \Pr\inbrak{Y_1 = x_1}} \Pr \inbrak{ Y_2 = x_2 }  \\
									 	&= \Pr \inbrak{Y_1 + Y_2 \ge t }
	\end{align*}
\end{proof}

\begin{corollary}
    If $X$ is a Poisson Binomial with parameters $p_1,\ldots,p_t,p_{t+1}$, and $Y$ is a Poisson Binomial with parameters $p_1,\ldots,p_t$, then
    \begin{equation}
        X \ge_{\mathrm{st}} Y
    \end{equation}
\end{corollary}

\section{Alternative Objective Function: Quality Model}
\label{sec:quality}

Our main model considers discrete types of producers (``honest'' and ``dishonest''). In addition to selecting honest producers, voters may be interested in electing block producers who will offer the highest \emph{performance} (e.g. transaction throughput). In this section, we outline an alternative model where producers vary continuously based on their ``quality'' (e.g. a metric of their computing performance, uptime and network latency). As before, voters receive a noisy signal about each producer's quality, and the voters use approval voting to elect a committee.

\begin{definition}[Continuous quality model]
Producer $j$ has an (unknown) quality, $q_j \sim \cQ$ for some (known) quality distribution $\cQ$.
As before, suppose voter $i$ receives a quality estimate $\hat{q}_{ij} = q_j + \eps_{ij}$, where $\eps_{ij} \sim N(0,\sigma_i)$.
\end{definition}

As above, voter $i$ will rank the block producers in order of of the signals $\inset{ q_{ij} }_j$, 
and may employ a \emph{Threshold} or \emph{Cardinality} voting strategy to determine how many block producers they choose to elect.

When focusing on honesty, we considered a one-shot game, since a single dishonest committee can potentially wreak havoc on the system, whereas a single low performing committee may only have a small effect on the overall popularity and utility of the system as a whole.

Thus in the quality model, it would make more sense to consider a \emph{multi-round} game where voters earn \emph{rewards} and receive \emph{feedback} at each round. When a set of block producers is elected, the voters can observe their throughput and latency, and thus get feedback about the quality of the committee.  We can consider different levels of granularity regarding the feedback received by the voters:
\begin{itemize}
	\item
	    \textbf{Individual feedback:}
		The quality of \emph{each} of the $c$ block producers that were elected to the committee
	\item
	    \textbf{Average feedback:}
		The \emph{average} quality of the $c$ block producers that were elected to the committee
\end{itemize}

We can also consider different types of voter rewards, which model the benefit they receive from higher throughput and and lower latency among the block-producer committee.
\begin{itemize}
	\item
	    \textbf{Average rewards:}
		The \emph{average} quality of the $c$ block producers that were elected to the committee
	\item
	    \textbf{Weakest-link rewards:}
		The \emph{minimum} quality of the $c$ block producers that were elected to the committee
\end{itemize}

Once we specify the exact type of feedback and rewards, we can ask how should voters behave in order to maximize their rewards. If a single voter could unilaterally specify the entire committee, the problem would fall in the class of \emph{combinatorial} multi-armed bandit problems (CMAB) \citep{CL12}.

Combinatorial bandit problems have been reasonably well studied \cite{CL12,CWY13,CTPL15,CHLL16,AA18,RM20} under two different feedback models (i) semi-bandit feedback, where the voters learn the quality of each individual producer in the committee and (ii) (full) bandit feedback, where the voters learn the average quality of the committee. There would be two main differences between our own model and traditional combinatorial multi-armed bandit problems:

First, most CMAB papers assume there is a single player who unilaterally selects which bandits to play (i.e., which candidates to elect).  In our setting, we have multiple voters, who vote independently, and the committee is chosen based on the outcome of this vote.

Second, most CMAB papers assume the player starts with no information about the underlying bandits (i.e., the voters receive no signals).  If voters start with no information about the committee (and all committee feedback is public), then for practical purposes, there is essentially only one voter, and the problem completely becomes a CMAB problem.

Given the difficulty of finding the optimal voting strategy in the static model, it will like be intractable to find the voting strategy that optimally combines with the CMAB exploration-exploitation strategy. But this could nonetheless be an interesting direction for future work.

\section{Other Asymptotic Results}
\label{app:other_asymptotics}

\begin{figure}[H]
	\begin{center}
		\begin{tikzpicture}
			\node (A) {\includegraphics[width=.4\textwidth]{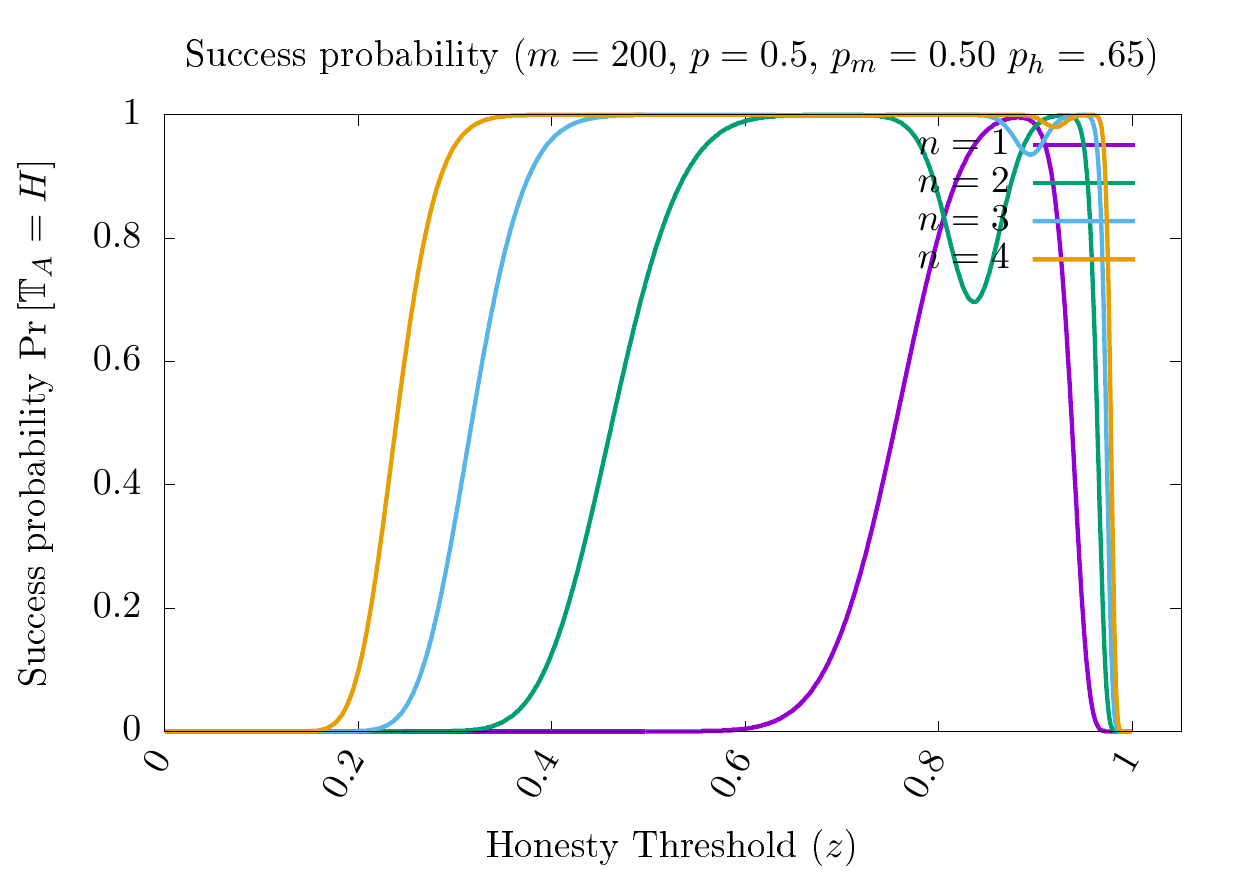}};
			\node (B) at ([xshift=1cm]A.east) [anchor=west] {\includegraphics[width=.4\textwidth]{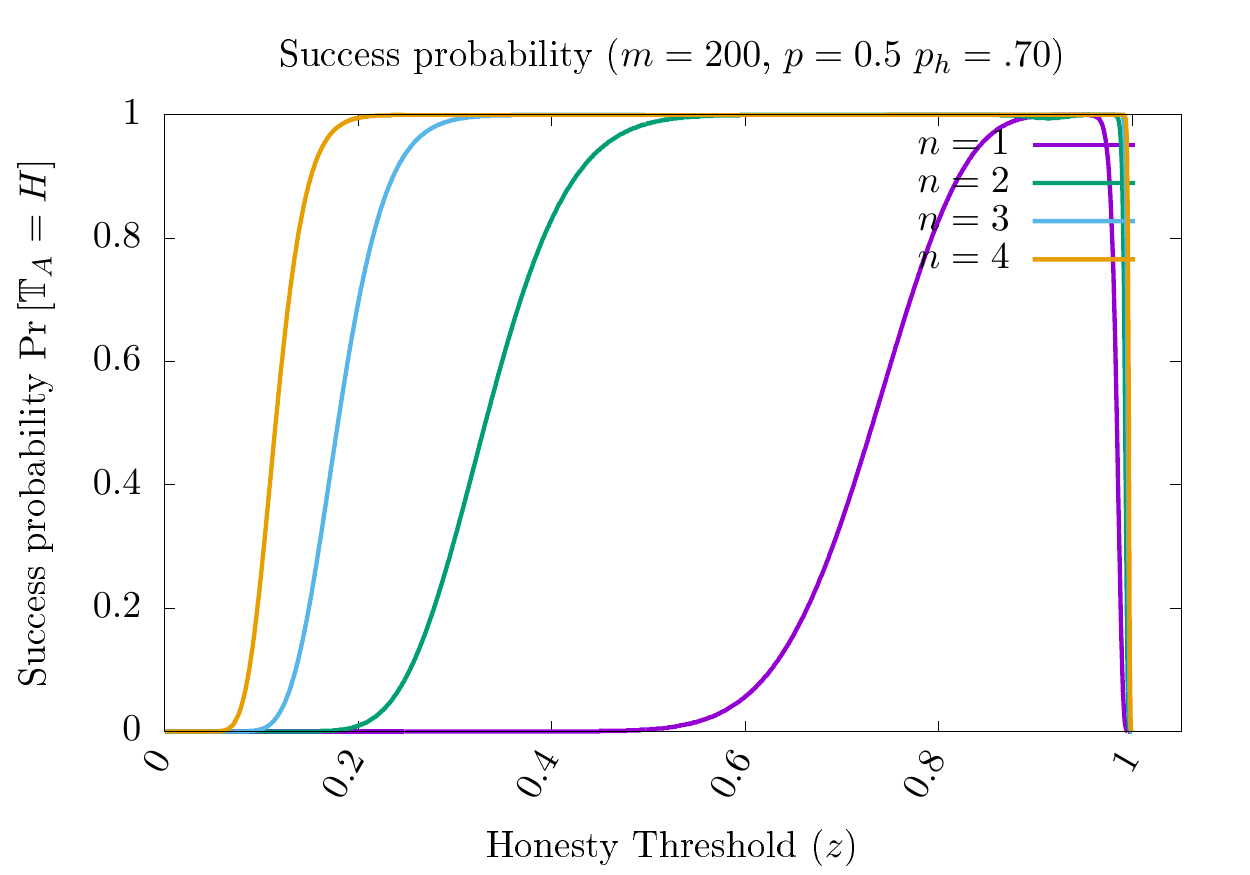}};
		\end{tikzpicture}
		\caption{Success probability as a function of voting threshold when signal informativeness is high. \label{fig:threshold_well_informed}}
	\end{center}
\end{figure}

Figure~\ref{fig:threshold_well_informed} shows that when the signal informativeness is high, that is, $p_h \gg p_m$, the success probability rapidly approaches 1 (even for a small number of voters), but if the threshold is too high $(z \approx 1)$, then the success probability drops to zero as all candidates receive 0 votes.

\begin{figure}[H]
	\begin{center}
	\includegraphics[width=.5\textwidth]{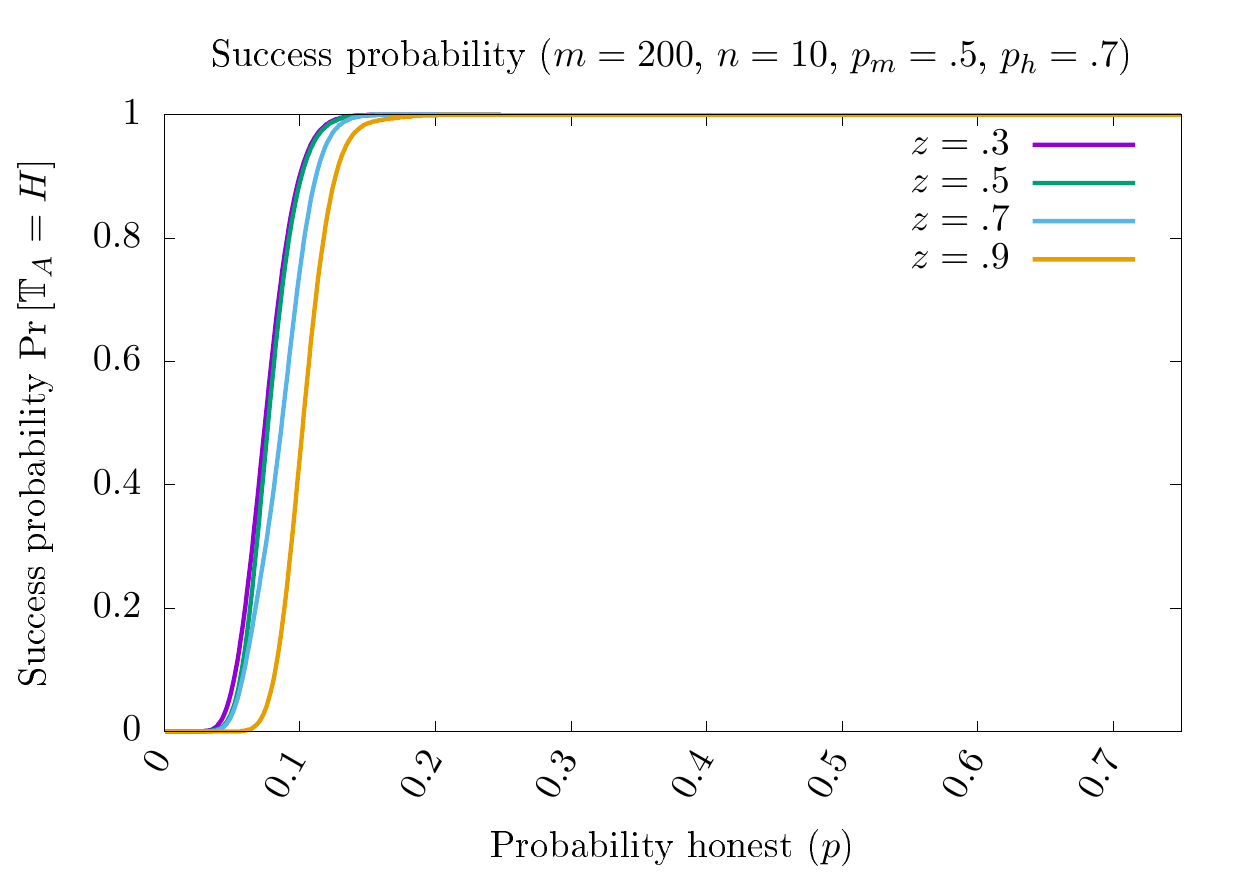}
		\caption{Success probability as a function of the prior. \label{fig:threshold_a_priori}}
	\end{center}
\end{figure}

Figure~\ref{fig:threshold_a_priori} shows that when the signal informativeness is high, as the \textit{a priori} probability that a producer is honest increases, then almost any threshold yields a nearly 100\% chance of electing an honest committee.

\section{Dependencies in Cardinal Voting}
\label{app:cardinal}

When voters follow a cardinal voting strategy, i.e., they vote for the $z$ candidates with the highest posterior probability of being honest, the analysis is Theorem~\ref{thm:success} no longer applies because the probabilities that each candidate receives a vote are no longer independent.  If you vote for candidate $j_1$, you are less likely to vote for candidate $j_2$, since you are only going to cast $z$ votes.  By contrast, when voters follow the threshold-voting strategy, the votes for different candidates \emph{are} independent.

Nevertheless, in the cardinal-voting setting, we can calculate the probabilities
$p_i^h$ (resp. $p_i^m$) denote the probability that voter $i$ 
votes for producer $j$ conditioned on $j$ being honest (resp. dishonest).

\begin{proposition}[Cardinal voting]
    \label{prop:cardinal}
    With notation as in Proposition~\ref{prop:threshold}, when voters follow the cardinal strategy (Definition~\ref{def:cardinal}) with cardinal, $z_i$, then
	\begin{align}
        p_i^h &= \sum_{a=0}^m{p^a(1-p)^{m-a}}  \int_{-\infty}^{\infty}{F_{Y_i(x)}f_{s_i|H}(x)\,dx},\\
        p_i^m &= \sum_{a=0}^m{p^a(1-p)^{m-a}}  \int_{-\infty}^{\infty}{F_{Y_i(x)}f_{s_i|M}(x)\,dx},
    \end{align}
    where 
    \begin{equation}
	\begin{aligned}
	    F_{Y_i}(x)&=\sum_{l=z_i}^{m}{\frac{(a!(m-a)!)^2}{z_i!(m-z_i)!}}\\
	    &\sum_{\substack{0\leq \lambda_1\leq z_i\\0\leq \lambda_2\leq m-z_i}}{\left[\binom{z_i}{\lambda_1}\inparen{F_{s_i|H}}^{\lambda_1}\inparen{F_{s_i|M}}^{z_i-\lambda_1}\binom{m-z_i}{\lambda_2}\inparen{1-F_{s_i|H}}^{\lambda_2}\inparen{1-F_{s_i|M}}^{m-z_i-\lambda_2}\right]}
	\end{aligned}
	\end{equation}
	and $f_{s_i|H},F_{s_i|H},f_{s_i|M},F_{s_i|M}$ are derived in Lemma~\ref{lem:posterior}.
\end{proposition}

Now, voter $i$ will vote for an honest candidate if the top-ranked honest candidate 
is higher than the top-ranked dishonest candidate.
Let $f_{s|H}$ (resp. $f_{s|M}$) denote the cdf of the signal, $s$, conditioned on a candidate 
being honest (resp. dishonest).  These cdfs are calculated explicitly in Equations~\ref{eqn:fsh} \& \ref{eqn:fsm}.

Suppose there are $a$ honest candidates and $b$ dishonest candidates,
in this setting, the CDF and pdf of the highest ranked honest candidate are
\begin{align}
	F_{H_{(a)}}(x) &= F_{s|H}^a \\
	f_{H_{(a)}}(x) &= a \inparen{ F_{s|H}(x) }^{a-1} \cdot f_{s|H}(x)
\end{align}
Similarly, the CDF and PDF of the highest ranked dishonest candidate are
\begin{align}
	F_{M_{(b)}}(x) &= F_{s|M}^b \\
	f_{M_{(b)}}(x) &= b \inparen{ F_{s|M}(x) }^{b-1} \cdot f_{s|M}(x)
\end{align}

Thus the probability that voter $i$ votes for an honest candidate is

\begin{align}
	\int_0^1 F_{M_{(b)}}(x) f_{H_{(a)}}(x) dx = a \int_0^1 F_{s|M}^b \cdot \inparen{ f_{s|H} }^{a-1} \cdot f_{s|H}(x) dx
\end{align}

Thus
\begin{align}
	p_i^h &= \sum_{a=1}^m \binom{n}{a} p^a(1-p)^{m-a} \inbrak{ \int_0^1 F_{s|M}^{m-a} \cdot \inparen{ f_{s|H} }^{a-1} \cdot f_{s|H}(x) dx } \\
	p_i^m &= \sum_{a=1}^m \binom{n}{a} p^a(1-p)^{m-a} \inbrak{ \int_0^1 F_{s|H}^{a} \cdot \inparen{ f_{s|M} }^{b-1} \cdot f_{s|M}(x) dx }
\end{align}

\end{document}